\documentclass[aip,amsmath,amssymb,reprint]{revtex4-2}
\usepackage{xcolor}
\usepackage{graphicx}
\usepackage{dcolumn}
\usepackage{bm}

\usepackage[utf8]{inputenc}
\usepackage[T1]{fontenc}
\usepackage{mathptmx}
\usepackage{etoolbox}

\usepackage{amsmath}
\usepackage{siunitx}
\usepackage{hyperref}
\usepackage{cleveref}
\crefname{appendix}{Appendix}{Appendices}
\Crefname{appendix}{Appendix}{Appendices}
\usepackage{url}
\usepackage{subcaption}

\begin{document}

\preprint{AIP/123-QED}

\title{Topological control of local electroneutrality in confined electrolytes} 



	\author{Marcelo Lozada-Cassou}
\email{marcelolcmx@ier.unam.mx} 

\affiliation{ Instituto de Energías Renovables, Universidad Nacional Autónoma de México (U.N.A.M.), Temixco, Morelos 62580, Mexico}


\date{\today}

\begin{abstract}
	
	Topology governs finite-size violations of local electroneutrality in confined
	electrolytes. Within Poisson--Boltzmann theory, we show that this topological
	control gives rise to a universal hierarchy of deviations in spherical,
	cylindrical, and planar confinement. We quantify this effect through an
	electroneutrality deviation ratio that captures the global electrostatic
	constraints associated with compactness and boundary multiplicity in
	$\Omega_{\mathrm{sph}}\simeq S^{2}\times[0,\delta]$,
	$\Omega_{\mathrm{cyl}}\simeq S^{1}\times\mathbb{R}\times[0,\delta]$, and
	$\Omega_{\mathrm{slit}}\simeq\mathbb{R}^{2}\times[0,\delta]$.
	Although local electroneutrality is asymptotically restored in the limit of
	infinite cavity size, finite-size deviations follow a robust topological
	hierarchy, being strongest in spherical cavities, weaker in cylindrical
	confinement, and weakest in planar slits. These results prove that topology is the organizing principle underlying confinement-induced charge redistribution and that violations of local	electroneutrality constitute a general electrostatic constraint governing overcharging, charge reversal, and long-range charge correlations. More fundamentally, they demonstrate that changing the topology modifies the global electrostatic constraints without altering the local Poisson--Boltzmann equations. As a counterintuitive manifestation of nonlinear confinement in a point-ion model, we report the existence of inside confinement charge reversal (ICCR) in charged hollow cylindrical and spherical nanoparticles. Since physically consistent, more general electrolyte theories must recover the Poisson--Boltzmann description in the appropriate limiting case, the present results establish a benchmark against which topological effects should be assessed beyond the Poisson--Boltzmann description.
	\end{abstract}


\maketitle 

\section{Introduction}\label{introduction}

Electrostatic interactions in confined electrolyte systems have been the subject of sustained interest due to their relevance in energy, colloidal science, soft matter, and biological physics~\cite{Kjellander-two-plates-1988,Evans-Wennerstrom-1999,Chmiola_2006,Supercapacitors-Book-2013,Lamperski-2014,Bohinc-2018,Wennerstrom_cells_2022,Feng-nanopores-topology-2023}. When charged fluids are confined within finite domains such as nanopores, nanochannels, or hollow nanoparticles, the presence of boundaries modifies both the structure of the electric double layer and the associated thermodynamic properties~\cite{bezadhi-encapsulation-2016,Boda-JCP-2026,Doan-Nguyen-nanoparticles-2023}. Deviations from local electroneutrality have been reported in confined geometries, highlighting the limitations of bulk-based intuitions in nanoscale systems~\cite{Lozada_1984,Lozada1996,Lozada-Cassou-PRL1996,Luo-electroneutrality-nature-2015,Levin_electroneutrality-2016,Levy-electroneutrality-2020,Levy-electroneutrality-PRE-2021,Rudi-Podgornik2023_JCIS,Rudi-French2010_RMP}. In particular, early gene-therapy approaches based on the self-assembly of DNA–cationic liposome complexes provided experimental examples of confined charged systems in which electrostatic correlations and membrane-mediated charge compensation can be interpreted as violations of local electroneutrality~\cite{Rädler-cationic-liposome-1997,Gelbart-Phys-Today-2000,Safinya-review-liposome-2001,Odriozola-PRL-2006,Odriozola_2006,odriozola-confined-macroions-2009}. Although these approaches have largely been superseded by more efficient modern gene-delivery strategies~\cite{Youssef-Review-cancer-2025,Geng-Viral-gene-therapy-2025}, confinement in hollow nanostructures remains central to a wide variety of biological and technological systems, including vesicles, liposomal carriers, gene-delivery nanoparticles, and nanoreactors.

A simple yet powerful theoretical framework to study the above kind of systems is the Poisson-Boltzmann (PB) theory~\cite{Verwey_TheoryStabilityLyophobicColloids_1948,McQuarrie_StatMech,Andelman-2006,Adrian-JML-2023,Boda-confinement-2024,Lozada-Cassou_JML-2025,Boda-JCP-2026}. While the validity of PB theory is restricted to low charges and electrostatic potentials on the shells of the hollow nanoparticles and to low electrolyte's concentrations, more sophisticated theories and computer simulations must reduce to the PB description in these limits~\cite{Lozada_1982,Degreve_1993,Degreve_1995,Gonzalez_2018,Boda-confinement-2024}. Moreover, both the PB analytical tractability and physical transparency are fundamental for understanding the basic mechanisms that govern electrostatic interactions in soft matter and electrochemical systems.

Among the interesting electrode-electrolyte interfacial phenomena are the charge reversal and overcharging of the effective electrode's charge. Traditionally these phenomena have been associated with excluded-volume and correlation effects arising from finite ionic or colloidal particle size, like in the inhomogeneous electrical double layer (EDL) of the restricted primitive model (RPM)~\cite{Kjellander-charge-inversion-1998,Deserno-2001,Lyklema-charge-reversal-2006} and primitive model (PM)~\cite{Jimenez_2004_Feb,gonzalez-Advances-2019,Gonzalez-overcharging-cyl-macroions-2022,Zeng-charge-reversal-cancer-2022,Samyabrata-overcharging-2024}. Remarkably, confinement overcharging (CO) and confinement charge reversal (CCR) have recently been reported even within a point-ion Poisson-Boltzmann description, in hollow nanoparticles immersed in a point ion electrolyte. These phenomena are exclusively associated with confinement and charge correlation between the inside and outside electrolyte, since no ionic size is considered beyond the Stern correction~\cite{Verwey_TheoryStabilityLyophobicColloids_1948}. 
These confinement-induced CO and CCR phenomena occur on the outside surface of the hollow nanoparticles shells. In the present work we show that analogous phenomena also occur on the inner surface, leading to what may be termed inside confinement charge reversal (ICCR). We demonstrate that all these phenomena originate from violations of the local electroneutrality condition (VLEC), which emerge naturally in finite confined systems; thus, VLEC in confined electrolytes can be understood as a direct consequence of these global constraints.

By analyzing spherical, cylindrical, and planar geometries within Poisson--Boltzmann theory, we investigate how the topology of confined domains governs finite-size violations of local electroneutrality and the resulting electrostatic response. Although local electroneutrality is recovered in the macroscopic limit, finite confined systems obey different global electrostatic constraints that depend on the topology of the confining domain. The central objective of this work is to demonstrate that topology constitutes the organizing principle governing finite-size violations of local electroneutrality in confined electrolytes.

Within Poisson--Boltzmann theory, electrostatic properties are typically determined by local differential equations supplemented by boundary conditions. However, confinement introduces global constraints that cannot be captured solely by local considerations. In particular, Gauss' law imposes integral relations that couple the charge distribution to the topology of the confining domain. As a consequence, systems governed by identical local electrostatic equations may exhibit qualitatively different behavior depending on whether the domain is compact or non-compact, and on the topology of the confining boundaries.~\cite{Nash_TopGeoPhys}.

\textit{Local electroneutrality is not a fundamental constraint in finite confined electrolytes. Its violation is governed by the topology of the confining domain, resulting in a correlation between the electrolyte inside and outside the cavity.}

In Section \ref{theory} we present a simple model for hollow nanoparticles of different geometries and briefly outline the linearized Poisson--Boltzmann solution for spherical, cylindrical, and planar nanocavities. In Sections \ref{results} and  \ref{discussion} we present our results and discuss their implications, respectively. The conclusions are presented in Section \ref{conclusions}.

\section{Theory}{\label{theory}
	
	\subsection{Geometry and topology}\label{topology}
Confinement profoundly modifies the electrostatic and thermodynamic behavior of electrolyte systems. When charged fluids are restricted to finite domains such as nanopores, hollow nanoparticles, or slit geometries, the presence of boundaries induces charge redistribution and deviations from local electroneutrality~\cite{Lozada_1984,Rudi-Parsegian-PRL-1998,Rudi-French2010_RMP,Rudi-Naji2013_Perspective,Rudi-Naji2014_SoftMatter,Levin_electroneutrality-2016,Levy-electroneutrality-2020,Levy-electroneutrality-PRE-2021,Green-electroneutrality-JCP-2021,Rudi-Podgornik2023_JCIS}. These effects are typically described in terms of geometric properties such as curvature or confinement length scales. However, despite extensive studies of confined electrolytes, a unifying principle that explains how global structural features of the domain control these phenomena remains lacking. Previous studies have reported confinement-induced charge redistribution~\cite{Adrian-JML-2023,Lozada-Cassou_JML-2025} and symmetry breaking~\cite{Lozada-Cassou_Symmetry_breaking_2025} in specific geometries.

In this article, we show that the finite-size violation of local electroneutrality in confined electrolytes is controlled by the topology of the confining domain. By analyzing spherical, cylindrical, and planar geometries within Poisson--Boltzmann theory~\cite{Verwey_TheoryStabilityLyophobicColloids_1948,Andelman-2006}, we establish a universal hierarchy of electroneutrality deviations determined by the compactness and connectivity of the domain. These results demonstrate that confinement-induced charge redistribution arises from global electrostatic constraints imposed by topology, rather than from local geometric features. Consequently, confinement-induced electrostatic, structural, transport, and thermodynamic properties are fundamentally determined by topology, while the specific geometry provides only its particular realization.

Hollow charged nanoparticles can be naturally described as product manifolds of the form
\begin{equation}
	\Omega \simeq \mathcal{M} \times [0,\delta],
\end{equation}
where $\mathcal{M}$ denotes the mid-surface of the shell (e.g., $\mathcal{M}=S^{2}$ for a spherical shell, $\mathcal{M}=S^{1}\times[0,L]$ for a finite cylindrical shell, or $\mathcal{M}=S^{1}\times\mathbb{R}$ for an infinite cylinder), and $[0,\delta]$ represents the transverse thickness. Within this representation, curvature acts as a geometric refinement, whereas the existence and structure of global constraints are determined by the topology of $\mathcal{M}$ and its boundary components.

Planar slit geometries represent a limiting case in which $\mathcal{M}  \simeq \mathbb{R}^{2}$. Formally,
\begin{equation}
	\Omega_{\text{slit}} \simeq \mathbb{R}^{2} \times [0,\delta].
\end{equation}

Here the domain is non-compact in the lateral directions and possesses translational invariance parallel to the confining walls. The boundary consists of two disconnected planar components.

Because $\mathbb{R}^{2}$ is non-compact, global constraints associated with closed surfaces are absent. Electrostatic quantities are typically defined per unit area, and thermodynamic functions like capacitance or osmotic pressure become an intensive properties. In contrast to spherical confinement, no global enclosure condition couples distant regions of the interface.

\textit{Thus, planar, cylindrical, and spherical geometries belong to distinct topological classes characterized by the compactness and connectivity of their mid-surfaces.}

\subsection{Model}\label{model}

We model the system as hollow nanoshells with planar, cylindrical, or spherical geometry, immersed in a bulk electrolyte. The electrolyte consists of symmetric point ions suspended in a continuous dielectric medium with dielectric constant $\varepsilon_{\scriptscriptstyle{0}}\varepsilon$. The nanoparticles are represented as shells with finite wall thickness $\delta$, such that the inner and outer surfaces of the shell are located at $r=R$ and $r=R+\delta$, respectively, measured from the geometric center. The dielectric constant of the shell's walls is taken to be equal to that of the bulk electrolyte, to avoid image potentials.

Each shell wall carries a constant surface charge density $\sigma_{\scriptscriptstyle{0}}$ on both inner and outer surfaces. Ion-ion interactions are modeled by the Coulomb potential, while ion-wall interactions are treated using a Stern layer correction~\cite{Verwey_TheoryStabilityLyophobicColloids_1948}, which assigns a finite ionic diameter $a$. Consequently, the electric double layers (EDLs) are restricted to the regions $0 \leq r \leq R - a/2$ and $R + \delta + a/2 \leq r < \infty$. See \cref{fig:geometry} for a schematic representation.

\begin{figure}[!htbp]
		\includegraphics[width=\linewidth]
		{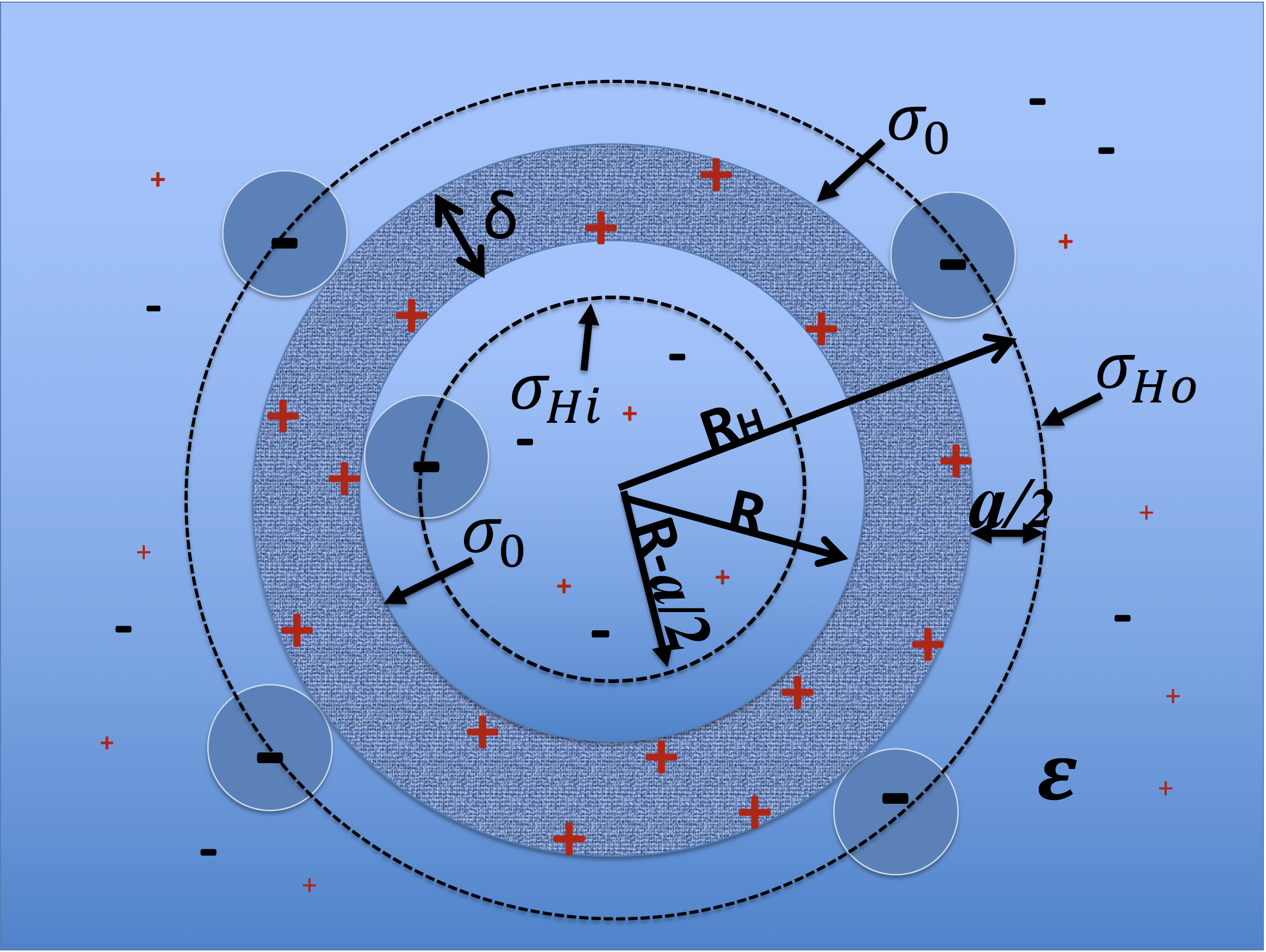}
	\caption{Schematic representation of a hollow nanoparticle showing the inner and outer fixed surface charge densities, $\sigma_0$, the induced inner and outer electrolyte charge densities, $\sigma_{\scriptscriptstyle{Hi}}$ and $\sigma_{\scriptscriptstyle{Ho}}$, the shell thickness $\delta$, and the Stern exclusion distance $a/2$. The notation is illustrated for the spherical geometry, but the same definitions apply to the cylindrical and planar geometries considered in this work. In the planar slit geometry, $R$ denotes one half of the slit width, measured from the midplane to either wall, and $R_{\scriptscriptstyle H}=R+\delta+a/2$ defines the outer Helmholtz plane.
}\label{fig:geometry}
\end{figure}

We define five spatial regions: $I$, $II$, $III$, $IV$, and $V$, corresponding to $0 \leq r \leq R - a/2$, $R - a/2 \leq r \leq R$, $R \leq r \leq R + d$, $R + d \leq r \leq R + d + a/2$, and $R + d + a/2 \leq r < \infty$, respectively. Regions $I$ and $V$ contain mobile ions and are governed by the Poisson equation,

\subsection{Poisson-Boltzmann equation}\label{PB-equation}

 The electrolyte's charge distribution, $\rho_{\scriptscriptstyle{el}}(r)$ in the regions $I$ and $V$ inside and outside the nanoparticles, respectively, is given by the Poisson equation~\cite{Jackson_2001}

\begin{equation}
	\nabla^2 \psi (r) = -\frac{1}{\varepsilon_{\scriptscriptstyle{0}}\varepsilon}\rho_{\scriptscriptstyle{el}}(r),
	\label{Poisson-equation}
\end{equation}

\noindent while in regions $II$--$IV$ (devoid of ions), the electrostatic potential satisfies Laplace's equation:

\begin{equation}
	\nabla^2 \psi (r) = 0.
	\label{Laplace-equation}
\end{equation}

\noindent Here, $r$ is measured radially from the geometrical center of the hollow nanoparticles (see \cref{fig:geometry})

If in \cref{Poisson-equation}, the local charge density is taken to be given by the Boltzmann distribution for point ions:

\begin{equation}
	\rho_{\scriptscriptstyle{el}}(r) = \sum_{i=1}^n e z_i \rho_{i0} \exp\left(-\beta e z_i \psi(r)\right),
	\label{Rho_elx2}
\end{equation}

\noindent where $\psi(r)$ is the mean electrostatic potential (MEP), $\rho_{i0}$ is the bulk concentration of ion species $i$, $z_i$, its valence, $e$ the elementary charge, $n$ the number of ionic species, and $\beta = 1/(kT)$. $k$ is the Boltzmann constant and $T$ is the system's temperature.

Substituting \cref{Rho_elx2} into \cref{Poisson-equation} gives the Poisson--Boltzmann (PB) equation. In the linear regime, where $\beta e z_i \psi \ll 1$, we obtain the linearized Poisson--Boltzmann (LPB) equation:

\begin{equation}
	\nabla^2 \psi(r) = \kappa^2 \psi(r),
	\label{eq:LPB}
\end{equation}

\noindent with

\begin{equation}
	\kappa = \sqrt{\frac{2 e^2 z^2 \rho_0}{\varepsilon_{\scriptscriptstyle{0}}\varepsilon k T}}=\frac{1}{\lambda_{\scriptscriptstyle{D}}},
	\label{Ec.kappa}
\end{equation}

\noindent where $\rho_0$ is the bulk concentration of each species in a symmetric $z:z$ electrolyte, and $\lambda_{\scriptscriptstyle{D}}$ denotes the Debye screening length~\cite{Hansen-book-2013}.

The electric field is obtained from the gradient of the potential,

\begin{equation}
	E(r) = -\nabla \psi(r) = \frac{1}{\varepsilon_{\scriptscriptstyle{0}}\varepsilon} \sigma(r),
	\label{electric-field general equation}
\end{equation}

\noindent where vector notation is omitted due to the azimuthal symmetry of the geometries considered (see \cref{fig:geometry}). 

The mean electrostatic potentials, $\psi(r)$, and surface charge densities, $\sigma(r)$,  at the interfaces of the hollow nanocavities are, 

\begin{itemize}
	\item $\phi_{\scriptscriptstyle{H}}$ and $\sigma_{\scriptscriptstyle{Ho}}$ at $r =R_{\scriptscriptstyle H} \equiv R + \delta + a/2$,
	\item $\phi_{\scriptscriptstyle{0}}$ and $\sigma_{\scriptscriptstyle{0}}$ at $r = R + \delta$,
	\item $\psi_{\scriptscriptstyle{0}}$ and $\sigma_{\scriptscriptstyle{0}}$ at $r = R$,
	\item $\psi_{\scriptscriptstyle{Hi}}$ and $\sigma_{\scriptscriptstyle{Hi}}$ at $r = R - a/2$,
	\item $\psi_{\scriptscriptstyle{d}}$ and zero charge at $r = 0$,
\end{itemize}

\noindent where $\sigma_{0}$ is the given fixed, surface charge density at both sides of  nanocavities' walls, while $\sigma_{\mathrm{Hi}}$ and $\sigma_{\mathrm{Ho}}$ are the induced surface charge densities on the electrolyte, inside and outside the nanocavity, respectively, located at $r=(R-a/2)$ and at $r=R+\delta+a/2$, in accordance with the Stern correction (see \cref{fig:geometry}).

To analytically solve the second order differential \cref{Laplace-equation,eq:LPB}, we will impose the following boundary condition: at $r=R$ and $r=R+\delta$ we assume the surface charge density on the hollow nanoparticles' wall to be both equal to a constant $\sigma_0$, and $\lim_{r \to \infty} \psi(r) = 0$ and $\lim_{r \to \infty} E(r) = 0$, which implies $\lim_{r \to \infty} \sigma(r) = 0$. 

These equations, together with the boundary conditions in accordance with the classical electrodynamics theory~\cite{Jackson_2001}, define the electrostatic problem in the LPB framework, whose analytical solution will be used to compute the mean electrostatic potential, and electric field profiles for the different hollow nanoparticles geometries. The resultant mean electrostatic potential, $\psi(r)$ and electric field $E(r)$, $\forall \quad 0 \leq r \leq\infty$, and as a function of $\sigma_{0}, R, \delta$ are given in Appendix~\ref{appendix-A}.

$\psi(r)$ and $\sigma(r)$ at all the nanoparticles' boundaries depend on $R$ and $\delta$. Hence, hereinafter, particularly for the induced inner and outer surface charge densities, we may explicitly express them as $\sigma_{\mathrm{Hi}}(R,\delta)$ and  $\sigma_{\mathrm{Ho}}(R,\delta)$. We will make this dependence explicit, when necessary. Additionally, hereinafter we may also use the subscripts $in$ and $out$ instead of $Hi$ and $Ho$, respectively to refer to these induced charges as  $\sigma_{\mathrm{in}}(R,\delta)$ and  $\sigma_{\mathrm{out}}(R,\delta)$ or 

The positive nanoparticle-ion, $g_{\scriptscriptstyle{\gamma +}}(r)$, and negative nanoparticle-ion, $g_{\scriptscriptstyle{\gamma -}}(r)$, inhomogeneous distribution functions inside and outside the nanoparticles, within the LPB equation, are given by

\begin{equation}
	\begin{split}
		&g_{\scriptscriptstyle{\gamma +}}(r) = \exp(-\,e\,z\,\beta\psi(r))\\& \quad r\in[0,(R-a/2)]\cup [(R+\delta+a/2),\infty),
		\label{Ec.gmas}
	\end{split}
\end{equation}

\noindent and

\begin{equation}
	\begin{split}
		&g_{\scriptscriptstyle{\gamma -}}(r) =\exp(\,e\,z\,\beta\psi(r))\\& \quad  r\in[0,(R-a/2)]\cup [(R+\delta +a/2),\infty),
		\label{Ec.gmenos}
	\end{split}
\end{equation}

\noindent where the subscript $\gamma$ referes to the nanoparicle species, as discussed below.

However, in this article we will not present explicit calculations for the EDL structure, which can be calculated through the explicit analytical solutions of $\psi(r)$ given in Appendix~\ref{appendix-A}. Some specific results for $g_{\scriptscriptstyle{+}}(r)$, and negative, $g_{\scriptscriptstyle{-}}(r)$ inside and outside nanopores can be found in previous publications\cite{Adrian-JML-2023,Lozada-Cassou_Symmetry_breaking_2025}.

\subsection{Electroneutrality and confinement}\label{electroneutrality}

In this article, we will focus on the properties of the reduced induced charges and induced surface-charge densities, inside, $\eta_{\mathrm{in}}(R,\delta)$, $\Sigma_{\mathrm{in}}(R,\delta)$, and outside, $\eta_{\mathrm{out}}(R,\delta)$, $\Sigma_{\mathrm{out}}(R,\delta)$, the cavity of the hollow nanoparticles, respectively,  defined as

\begin{align}
	\Sigma_{\mathrm{in}}(R,\delta)=\frac{\sigma_{\mathrm{0}}+\sigma_{\mathrm{in}}(R,\delta)}{\sigma_{\mathrm{0}}}, \label{Sigma-in} \\
		\Sigma_{\mathrm{out}}(R,\delta)=\frac{\sigma_{\mathrm{out}}(R,\delta)-\sigma_{\mathrm{0}}}{\sigma_{\mathrm{0}}}, \label{Sigma-out}
\end{align}

\begin{align}
	\eta_{\mathrm{in}}(R,\delta)=\frac{Q_{\mathrm{0}}(R)+Q_{\mathrm{in}}(R,\delta)}{Q_{\mathrm{0}}(R)}, \label{Eta-in} \\
	\eta_{\mathrm{out}}(R,\delta)=\frac{Q_{\mathrm{out}}(R,\delta)-Q_{\mathrm{0}}(R+\delta)}{Q_{\mathrm{0}}(R+\delta)}. \label{Eta-out}
\end{align}

\noindent Where $Q_{\mathrm{0}}(R)$ is the given fixed charge at $r=R$, $Q_{\mathrm{0}}(R+\delta)$ at $r=(R+\delta)$, and $Q_{\mathrm{in}}(R,\delta) \equiv Q_{\mathrm{Hi}}(R,\delta)$ and $Q_{\mathrm{out}}(R,\delta) \equiv Q_{\mathrm{Ho}}(R,\delta)$ are the charges at $r=(R-a/2)$ and at $r=(R+\delta +a/2) \equiv R_{\mathrm{H}}$, respectively.

In terms of $\sigma_{\mathrm{in}}(R,\delta)$ and  $\sigma_{\mathrm{out}}(R,\delta)$, \cref{Eta-in,Eta-out} become

\begin{align}
	\eta_{\mathrm{in}}(R,\delta)=\frac{[R^{\gamma}\sigma_{\mathrm{0}}+(R-a/2)^{\gamma}\sigma_{\mathrm{in}}(R,\delta)]}{R^{\gamma}\sigma_{\mathrm{0}}}, \label{Eta-sig-in} \\
	\eta_{\mathrm{out}}(R,\delta)=\frac{[R^{\gamma}_{\mathrm{H}}\sigma_{\mathrm{out}}(R,\delta)-(R+\delta)^{\gamma}\sigma_{\mathrm{0}}]}{(R+\delta)^{\gamma}\sigma_{\mathrm{0}}}, \label{Eta-sig-out}
\end{align}
\noindent where \( \gamma = 0, 1, 2 \) for planar, cylindrical, and spherical geometries, respectively.

To explicitly calculate $\eta_{\mathrm{in}}(R,\delta)$, $\Sigma_{\mathrm{in}}(R,\delta)$, $\eta_{\mathrm{out}}(R,\delta)$, and $\Sigma_{\mathrm{out}}(R,\delta)$ we need the expressions for $\sigma_{\mathrm{in}}(R,\delta)\equiv\sigma_{\mathrm{Hi}}(R,\delta)$ and $\sigma_{\mathrm{out}}(R,\delta)\equiv\sigma_{\mathrm{Ho}}(R,\delta)$, in terms of the boundary surface charge density on the hollow nanoparticles walls, given in Appendix~\ref{appendix-A}.

For all three hollow nanoparticle geometries, the total charge balance is given by:
\begin{equation}\label{Electroneutrality-condition-general}
	\begin{split}
		Q_{\scriptscriptstyle{0}}(R) + Q_{\scriptscriptstyle{0}}(R+ \delta) + Q_{\scriptscriptstyle{Hi}}(R - a/2)\\
		+ Q_{\scriptscriptstyle{Ho}}(R +  \delta + a/2) = 0,
	\end{split}
\end{equation}
\noindent where \( Q_{\scriptscriptstyle{0}}(R) \) and \( Q_{\scriptscriptstyle{0}}(R+ \delta) \) are the fixed surface charges located at \( A_\gamma(R) \) and \( A_\gamma(R+ \delta) \), respectively, and \( Q_{\scriptscriptstyle{Hi}} \) and \( Q_{\scriptscriptstyle{Ho}} \) represent the induced charges in the inner and outer electric double layers (EDLs). These are defined as:
\begin{equation}\label{Induced-charge-in}
	\begin{split}
		Q_{\scriptscriptstyle{Hi}}(R - a/2) &= \int\limits_{\boldsymbol{\omega_{\scriptscriptstyle{in}}}} \rho_{\text{el}}(r) \, d\mathbf{V}\\ &= f_{\scriptscriptstyle{\gamma}} \int_{0}^{R - a/2} r^\gamma \rho_{\text{el}}(r) \, dr, 
	\end{split}
\end{equation}

\noindent and

\begin{equation}\label{Induced-charge-out}
	\begin{split}
		Q_{\scriptscriptstyle{Ho}}(R +  \delta + a/2) &= \int\limits_{\boldsymbol{\omega_{\scriptscriptstyle{out}}}} \rho_{\text{el}}(r) \, d\mathbf{V}=\\
		& f_{\scriptscriptstyle{\gamma}} \int_{R +  \delta + a/2}^{\infty} r^\gamma \rho_{\text{el}}(r) \, dr.
	\end{split}
\end{equation}

\noindent The integrals in \cref{Induced-charge-in,Induced-charge-out} are evaluated over the internal and external electrolyte volumes, \( \boldsymbol{\omega_{\scriptscriptstyle{in}}} \) and \( \boldsymbol{\omega_{\scriptscriptstyle{out}}} \), respectively. Here, \( f_{\scriptscriptstyle{\gamma}} \) is a geometry-dependent constant, where \( \gamma = 0, 1, 2 \).

From these expressions, the induced surface charge density profiles for \( r \leq R - a/2 \) and \( r \geq R +  \delta + a/2 \) are given by:
\begin{equation} \label{Induced-charge-density-in}
	\sigma_{\scriptscriptstyle{\gamma}}(r)= \frac{1}{r^\gamma} \int_0^r r^\gamma \rho_{\text{el}}(r) \, dr,
\end{equation}

\noindent and

\begin{equation}\label{Induced-charge-density-out}
	\sigma_{\scriptscriptstyle{\gamma}}(r)= -\frac{1}{r^\gamma} \int_r^\infty r^\gamma \rho_{\text{el}}(r) \, dr. 
\end{equation}

Substituting \cref{Induced-charge-density-in,Induced-charge-density-out} into \cref{Electroneutrality-condition-general} yields the general electroneutrality condition:

\begin{equation}\label{Electroneutrality-condition-general2}
	\begin{split}
		&R^\gamma \sigma_{\scriptscriptstyle{0}} + (R +  \delta)^\gamma \sigma_{\scriptscriptstyle{0}} + (R - a/2)^\gamma \sigma_{\scriptscriptstyle{\gamma}}(R - a/2)\\
		&\quad\quad= (R +  \delta + a/2)^\gamma \sigma_{\scriptscriptstyle{\gamma}}(R +  \delta + a/2).
	\end{split}
\end{equation}

\noindent This equation applies to all three geometries presented in Appendix~\ref{appendix-A} and allows for the analytical determination of \( \sigma_{\scriptscriptstyle{Hi}} \) when \(\sigma_{\scriptscriptstyle{Ho}} \) is known.

The terms \( \sigma_{\scriptscriptstyle{\gamma}}(R - a/2)/(\varepsilon_{\scriptscriptstyle{0}}\varepsilon) \) and \( \sigma_{\scriptscriptstyle{\gamma}}(R +  \delta + a/2)/(\varepsilon_{\scriptscriptstyle{0}}\varepsilon) \) correspond to the effective electric fields at \( r = R - a/2 \) and \( r = R +  \delta + a/2 \), respectively. Thus, \cref{Electroneutrality-condition-general2} implies an electric field balance:

\begin{equation}\label{Electrical-field-balance}
	\begin{split}
		&R^\gamma E_\gamma(R) \mathbf{e_r} + (R - a/2)^\gamma E_\gamma(R - a/2) \mathbf{e_r}\\
		& + (R +  \delta)^\gamma E_\gamma(R + d) \mathbf{e_r} \\
		&= (R +  \delta + a/2)^\gamma E_\gamma(R +  \delta + a/2) \mathbf{e_r},
	\end{split}
\end{equation}

\noindent where \( \mathbf{e_r} \) is the unit radial vector, and in agreement with \cref{Electroneutrality-condition-general2}, where $
\sigma_{\scriptscriptstyle{Hi}} \equiv \sigma_{\scriptscriptstyle{\gamma}}(R - a/2)$ and $\sigma_{\scriptscriptstyle{Ho}} \equiv \sigma_{\scriptscriptstyle{\gamma}}(R +  \delta + a/2).$

\vspace{\baselineskip}

It can be shown that:
\begin{equation}\label{Eq.Violation_Local_Elec_Cond}
	E_\gamma(R - a/2)\mathbf{e_r} + E_\gamma(R)\mathbf{e_r} \neq \vec{0}, \quad \text{for all finite  R}.
\end{equation}

\noindent Thus, since \( E_\gamma(R)=\sigma_0/(\varepsilon_0\varepsilon) \), the local electroneutrality condition is violated within the shell. This finite-size violation of local electroneutrality (VLEC) originates from the topology of the nanoparticles, which gives rise to ion--ion charge correlations between the electrolytes inside and outside the shell~\cite{Lozada-Cassou-PRE1997}. This phenomenon has been predicted by Poisson--Boltzmann and integral equation theories~\cite{Lozada_1984,Lozada1996,Lozada-Cassou-PRL1996,Yu_1997,Aguilar_2007,Levin_electroneutrality-2016}, and subsequently corroborated by density functional theory, computer simulations~\cite{Levy-electroneutrality-2020,Levy-electroneutrality-PRE-2021,Keshavarzi_2020}, and experiments~\cite{Cuvillier-Rondelez-1998,Luo-electroneutrality-nature-2015}.

Nonetheless, global electroneutrality is preserved via \cref{Electrical-field-balance}. While slit-shells may approach local neutrality for moderate \( R \), spherical and cylindrical shells require much larger \( R \) to do so. However, in the limit \( R \to \infty \),
\begin{equation}\label{electroneutrality-in}
	\lim_{R \to \infty} [E_\gamma(R - a/2) + E_\gamma(R)] = 0,
\end{equation}

\noindent and

\begin{equation}\label{electroneutrality-out}
	\lim_{R \to \infty} [E_\gamma(R + d + a/2) - E_\gamma(R + d)] = 0.
\end{equation}
Thus, the local electroneutrality condition is recovered asymptotically. In this limit, \cref{Electrical-field-balance} reduces to

\begin{equation}
	\begin{split}
		(R - a/2)^\gamma \sigma_0+R^\gamma \sigma_0 &+ (R + d)^\gamma \sigma_0=\\ 
		&(R + d + a/2)^\gamma \sigma_0,
	\end{split}
\end{equation}
implying that in this limit \textit{local electroneutrality is independently satisfied inside and outside the shell}.

However, it should be pointed out that achieving electroneutrality numerically, through \cref{Induced-charge-in,Induced-charge-out,electroneutrality-in,electroneutrality-out} is virtually impossible, for $\gamma=1,2$, due to the very large charge correlations imposed by the VLEC. Notwithstanding, this limit can be demonstrated analytically (see Appendix~\ref{appendix-B}).However, for the slit, $\gamma=0$, in general, electroneutrality can be numerically obtained.

 A nonzero value of $\eta_{\mathrm{in}}(R,\delta)$ and $\eta_{\mathrm{out}}(R,\delta)$ signals a violation of local electroneutrality, i.e., these quantities provide complementary measures of confinement-induced charge imbalance, while $\Sigma_{\mathrm{in}}(R,\delta)$ and $\Sigma_{\mathrm{out}}(R,\delta)$ characterize the corresponding effective induced surface charge densities at the inner and outer interfaces. 
 
 The electrostatic potential is determined self-consistently by the global charge distribution, which couples the interior and exterior regions of the confined domain through long-range Coulomb interactions~\cite{Lozada-Cassou-PRE1997}. As a result, the charge distribution inside the cavity cannot be determined independently of the exterior region, signaling correlations across the nanocavity walls, which share the same bulk chemical potential, $\mu^{bulk}$.

This definition enables a direct comparison of confinement-induced charge imbalance across geometries with different topological properties. To compare the overall charge imbalance produced by confinement across different topologies, we introduce the reduced global electroneutrality imbalance (GEI), $\eta_{\mathrm{T}}(R,\delta)$, defined as

\begin{equation}\label{Eq:Total-VLEC}
	\begin{split}
&\eta_{\mathrm{T}}(R,\delta)\equiv\\
&\frac{R^\gamma \sigma_{\scriptscriptstyle{0}} + (R +  \delta)^\gamma \sigma_{\scriptscriptstyle{0}} + (R - a/2)^\gamma \sigma_{\scriptscriptstyle{\gamma}}(R - a/2)\quad\quad}{R^\gamma \sigma_{\scriptscriptstyle{0}} + (R +  \delta)^\gamma \sigma_{\scriptscriptstyle{0}}}\\
&=\frac{(R +  \delta + a/2)^\gamma \sigma_{\scriptscriptstyle{\gamma}}(R +  \delta + a/2)}{R^\gamma \sigma_{\scriptscriptstyle{0}} + (R +  \delta)^\gamma \sigma_{\scriptscriptstyle{0}}}.
\end{split}
\end{equation}

\noindent See \cref{Electroneutrality-condition-general2}. Of course, the left-hand side and the right-hand side of \cref{Eq:Total-VLEC} are equal.

The reduced quantity $\eta_{\mathrm{T}}(R,\delta)$ measures the global electroneutrality imbalance (GEI) of the confined electrolyte. The corresponding reduced quantities $\eta_{\mathrm{in}}(R,\delta)$ and $\eta_{\mathrm{out}}(R,\delta)$ quantify the local electroneutrality imbalance inside the cavity and outside the hollow shell, respectively. The reduced quantities $\Sigma_{\mathrm{in}}(R,\delta)$ and $\Sigma_{\mathrm{out}}(R,\delta)$ characterize the corresponding local electroneutrality imbalance manifested through the inner and outer electrical double layers, respectively. Together, these five reduced observables provide a unified description of topology-controlled violations of local electroneutrality, linking the global imbalance to its corresponding local volumetric and interfacial manifestations.

\noindent The physical significance of $\Sigma_{\mathrm{out}}(R,\delta)$ extends throughout the external electrolyte. Once the topology-controlled local electroneutrality imbalance modifies the induced charge density at the outer shell surface, it determines the electric field
\[
E_{\scriptscriptstyle{\gamma}}(r,\delta)=\sigma_{\scriptscriptstyle{\gamma}}(r,\delta)/(\varepsilon_{\scriptscriptstyle{0}}\varepsilon),
\]
throughout the external electrolyte, i.e., for all $r>R+\delta+a/2$. The resulting statistical-mechanical correlations within the confined system, linking the electrolytes inside and outside the shell, are likewise transmitted throughout the external electrolyte and are reflected in the inhomogeneous nanoparticle--ion correlation functions, $g_{\scriptscriptstyle{\gamma j}}(r)$. Consequently, the influence of topology is not confined to the immediate vicinity of the shell, but propagates throughout the surrounding electrolyte, thereby affecting electrostatic interactions, osmotic forces, colloidal stability, electrophoretic mobility, and other structural, transport, and thermodynamic properties.

\section{Results}\label{results}

In this section, we present results for $\eta_{\mathrm{T}}(R,\delta)$, $\Sigma_{\mathrm{in}}(R,\delta)$, $\Sigma_{\mathrm{out}}(R,\delta)$, $\eta_{\mathrm{in}}(R,\delta)$ and $\eta_{\mathrm{out}}(R,\delta)$, as a function of the reduced cavity radius, $R/(a/2)$, for  some representative values of $\rho_0$, $\sigma_{0}$, and $\delta$. In all our calculations, we consider a positively charged shell immersed in an aqueous symmetric electrolyte ($1:1$ or $2:2$), with a relative dielectric constant $\varepsilon = 78.5$, temperature $T=298.15~K$, and ionic diameters, $a=4.25 \textup{\r{A}}$.

\Cref{fig:Eta2-d2-d200-sig0p005-s0p0005-rho0p1-rho0.001} shows the behavior of the GEI, $\eta_{\mathrm{T}}(R,\delta)$, as a function of the reduced cavity size, $R/(a/2)$, for planar, cylindrical, and spherical geometries, obtained from the analytical solutions of the linearized Poisson-Boltzmann \cref{eq:LPB}~\cite{Adrian-JML-2023,Lozada-Cassou_Symmetry_breaking_2025}. 

In all cases, $\eta_{\mathrm{T}}(R,\delta)$ decreases monotonically and satisfies
$\lim_{R\to\infty}\eta_{\mathrm{T}}(R,\delta)=0.5$,
corresponding to the separate compensation of the fixed charges on the inner and outer shell surfaces by their respective induced charges, and hence to the recovery of local electroneutrality on both sides of the hollow-shell wall (see Appendix~\ref{appendix-B}). However, at finite cavity size, the magnitude of the deviation exhibits a clear and systematic ordering across geometries. The strongest deviations occur in spherical confinement, followed by cylindrical and planar geometries, yielding the hierarchy
\begin{equation} \label{Eq:VLEC-ordering}
	\eta_{\mathrm{T,spherical}}(R,\delta) > \eta_{\mathrm{T,cylindrical}}(R,\delta) > \eta_{\mathrm{T,planar}}(R,\delta).
\end{equation}
This ordering is insensitive to variations in electrolyte concentration and valence, surface charge density, and shell-wall thickness, and reflects a structural property of the confining domain rather than local geometric features alone. Thence, although the magnitude of the GEI may change with ionic valence, electrolyte concentration and/or shell-wall thickness, and/or surface charge density, the topology-controlled confinement hierarchy persists. Therefore, the ordering sphere > cylinder > slit is a robust feature of confined electrolytes. Moreover, $\eta_{\mathrm{T}}(R,\delta)$ is invariant under any value of the shell's surface charge density, $\sigma_{\mathrm{0}}$.

	\begin{figure}[!htbp]
	\centering
	
	\begin{subfigure}[b]{0.48\columnwidth}
		\centering
		\includegraphics[width=\linewidth]
		{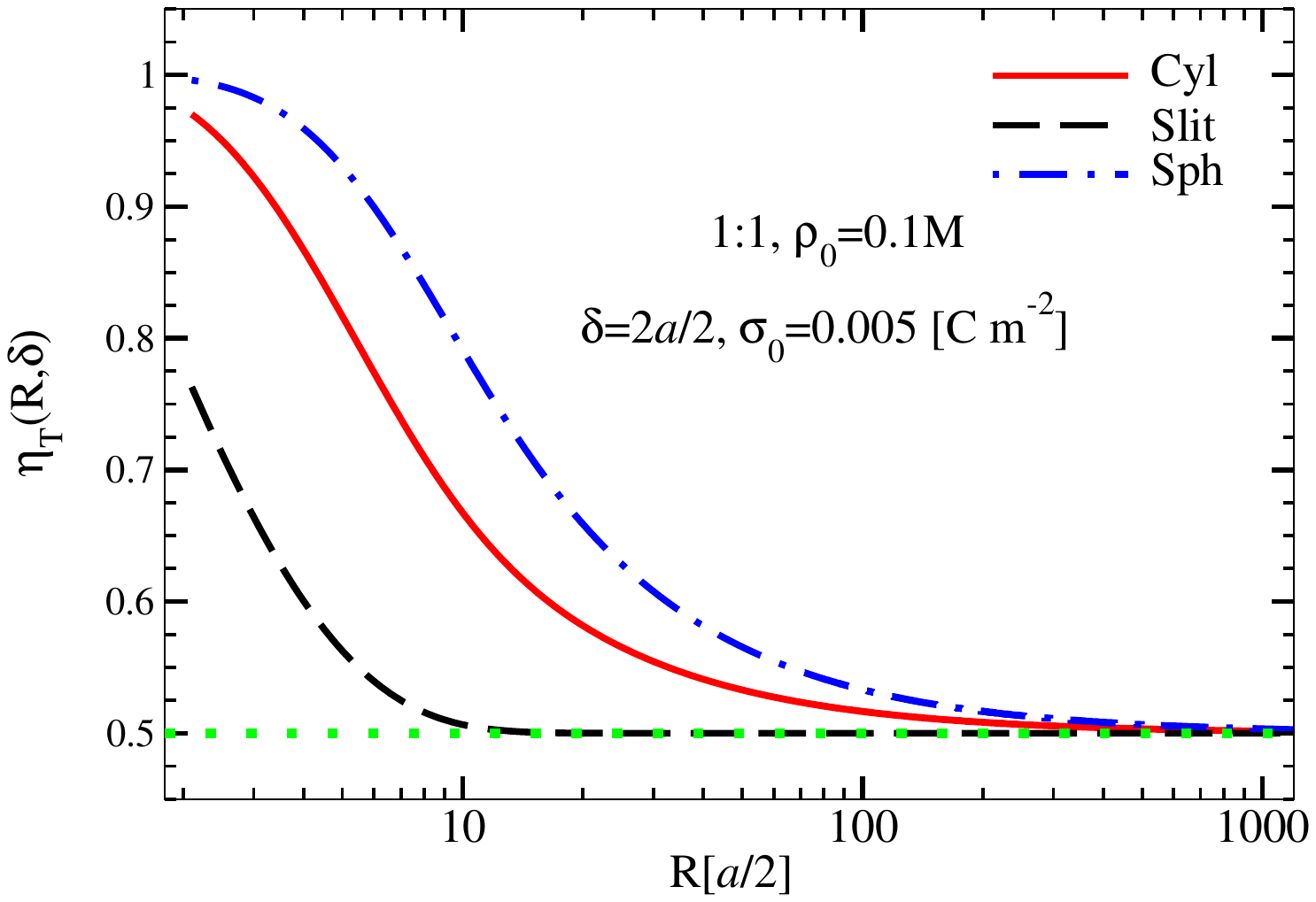}
		\caption{Thin nanocavities' walls}
		\label{fig:etaT-d2-s0p005-rho0p1-z1}
	\end{subfigure}
	\hfill
	\begin{subfigure}[b]{0.48\columnwidth}
		\centering
		\includegraphics[width=\linewidth]
		{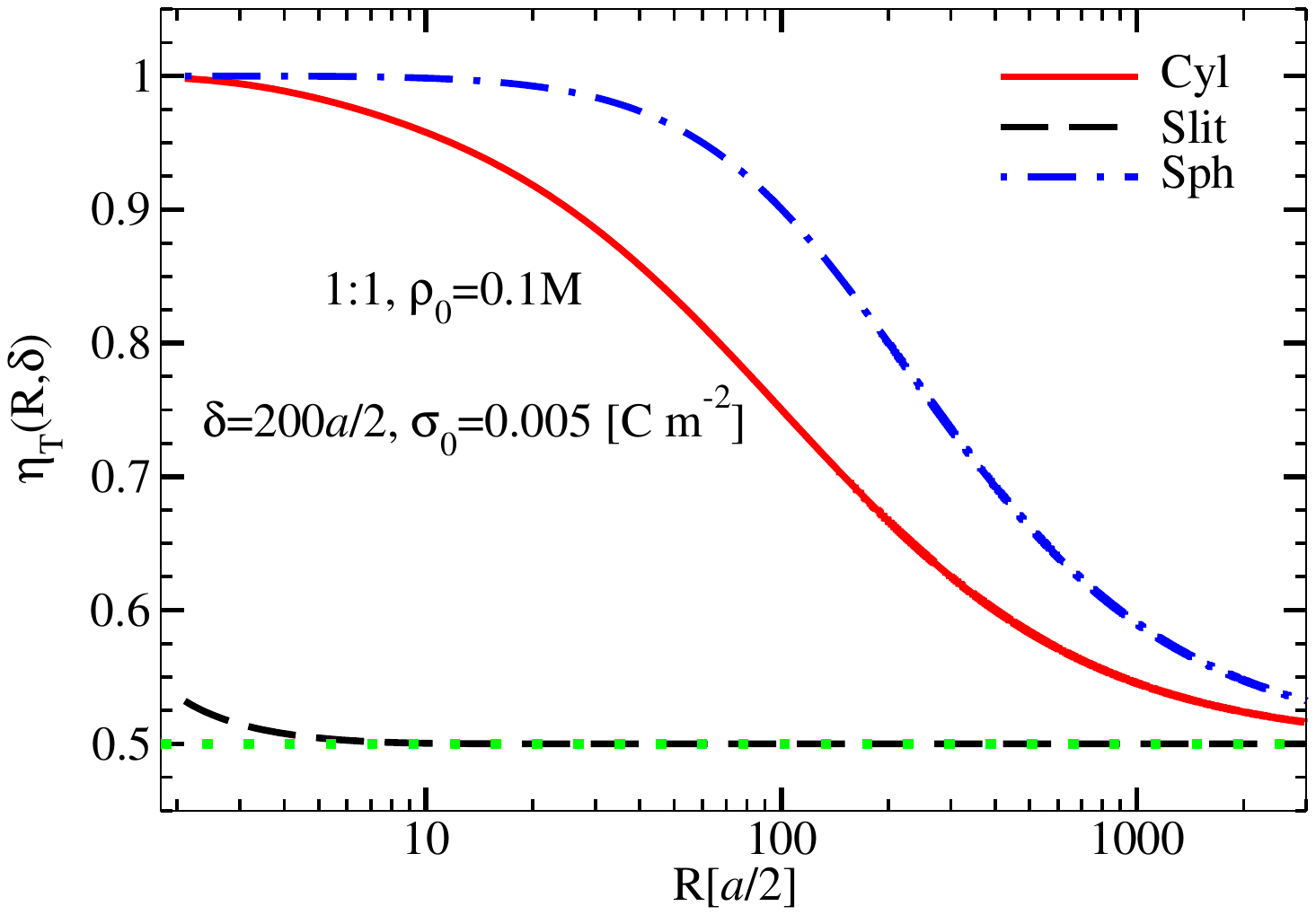}
		\caption{Thick nanocavities' walls}
		\label{fig:etaT-d200-s0p005-rho0p1-z1}
	\end{subfigure}
	
	\vspace{0.3cm}
	
	\begin{subfigure}[b]{0.48\columnwidth}
		\centering
		\includegraphics[width=\linewidth]
		{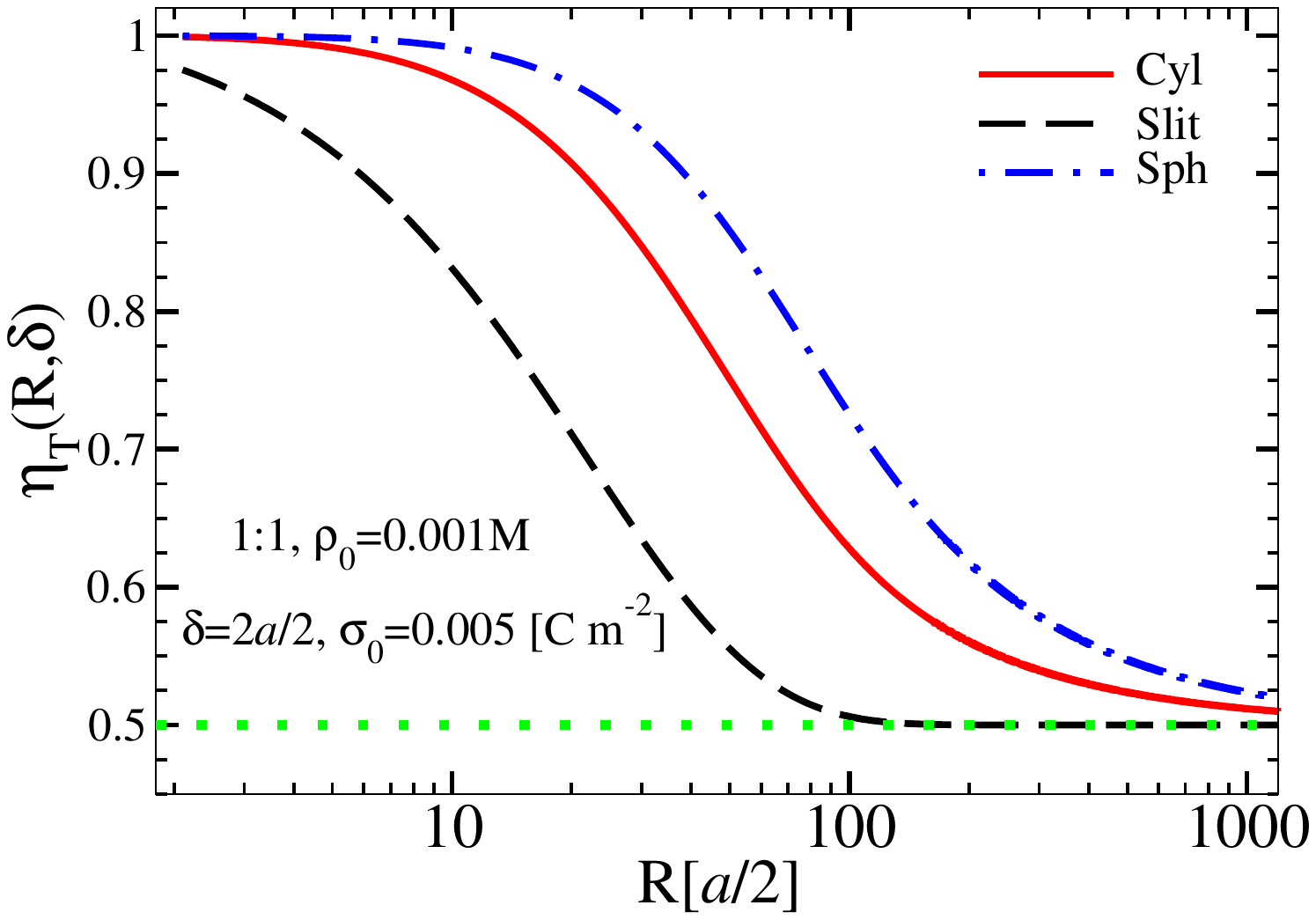}
		\caption{Very low salt concentration}
		\label{fig:EtaT-d2-s0p005-rho0p001-z1}
	\end{subfigure}
	\hfill
	\begin{subfigure}[b]{0.48\columnwidth}
		\centering
		\includegraphics[width=\linewidth]
		{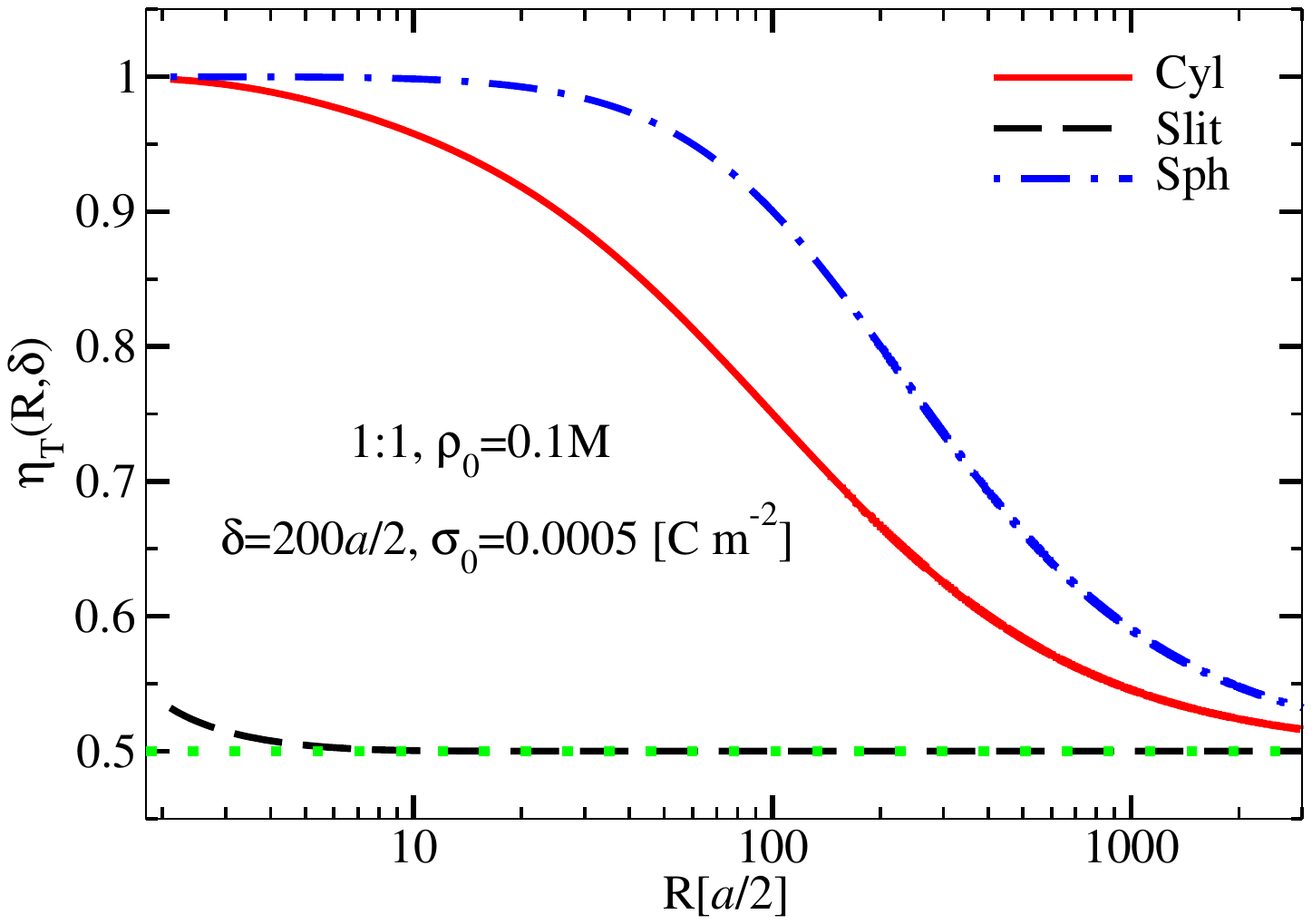}
		\caption{Lower surface charge density}
		\label{fig:EtaT-d200-s0p0005-rho0p1-z1}
	\end{subfigure}
	\caption{Reduced global electroneutrality imbalance (GEI), $\eta_{\mathrm{T}}(R,\delta)$, as a function of the reduced cavity radius $R/(a/2)$ (logarithmic scale) for electrolytes confined within planar slits ($\Omega_{\mathrm{slit}}\simeq \mathbb{R}^{2}\times[0,\delta]$), infinite cylindrical shells ($\Omega_{\mathrm{cyl}}\simeq S^{1}\times\mathbb{R}\times[0,\delta]$), and spherical shells ($\Omega_{\mathrm{sph}}\simeq S^{2}\times[0,\delta]$), for the model parameters indicated in each panel. Finite confined systems exhibit a robust topology-controlled hierarchy of the GEI, with spherical cavities displaying the largest imbalance, followed by cylindrical shells and planar slits. In the macroscopic limit, $\displaystyle\lim_{R\rightarrow\infty}\eta_{\mathrm{T}}(R,\delta)=0.5$, corresponding to equal compensation of the total fixed surface charge on the inner and outer shell surfaces by the corresponding induced charges. The horizontal dotted line denotes this asymptotic limit.
	}
	\label{fig:Eta2-d2-d200-sig0p005-s0p0005-rho0p1-rho0.001}
\end{figure}

	\begin{figure}[!htbp]
		\centering
		
		\begin{subfigure}[b]{0.48\columnwidth}
			\centering
			\includegraphics[width=\linewidth]
			{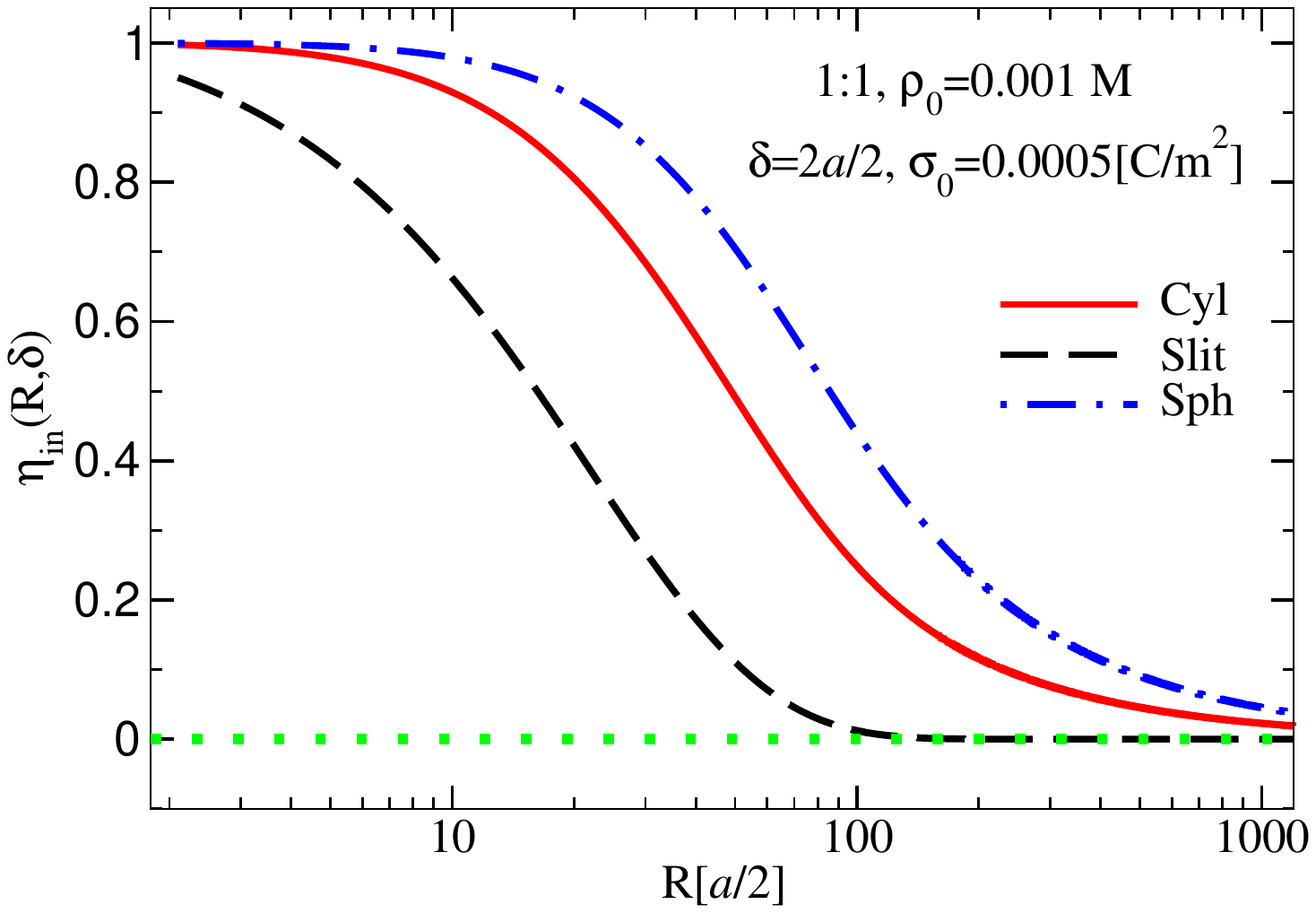}
			\caption{$\eta_{\mathrm{in}}(R,\delta)$}
			\label{fig:EtaIn-d2-s0p0005-rho0p001}
		\end{subfigure}
		\hfill
		\begin{subfigure}[b]{0.48\columnwidth}
			\centering
			\includegraphics[width=\linewidth]
			{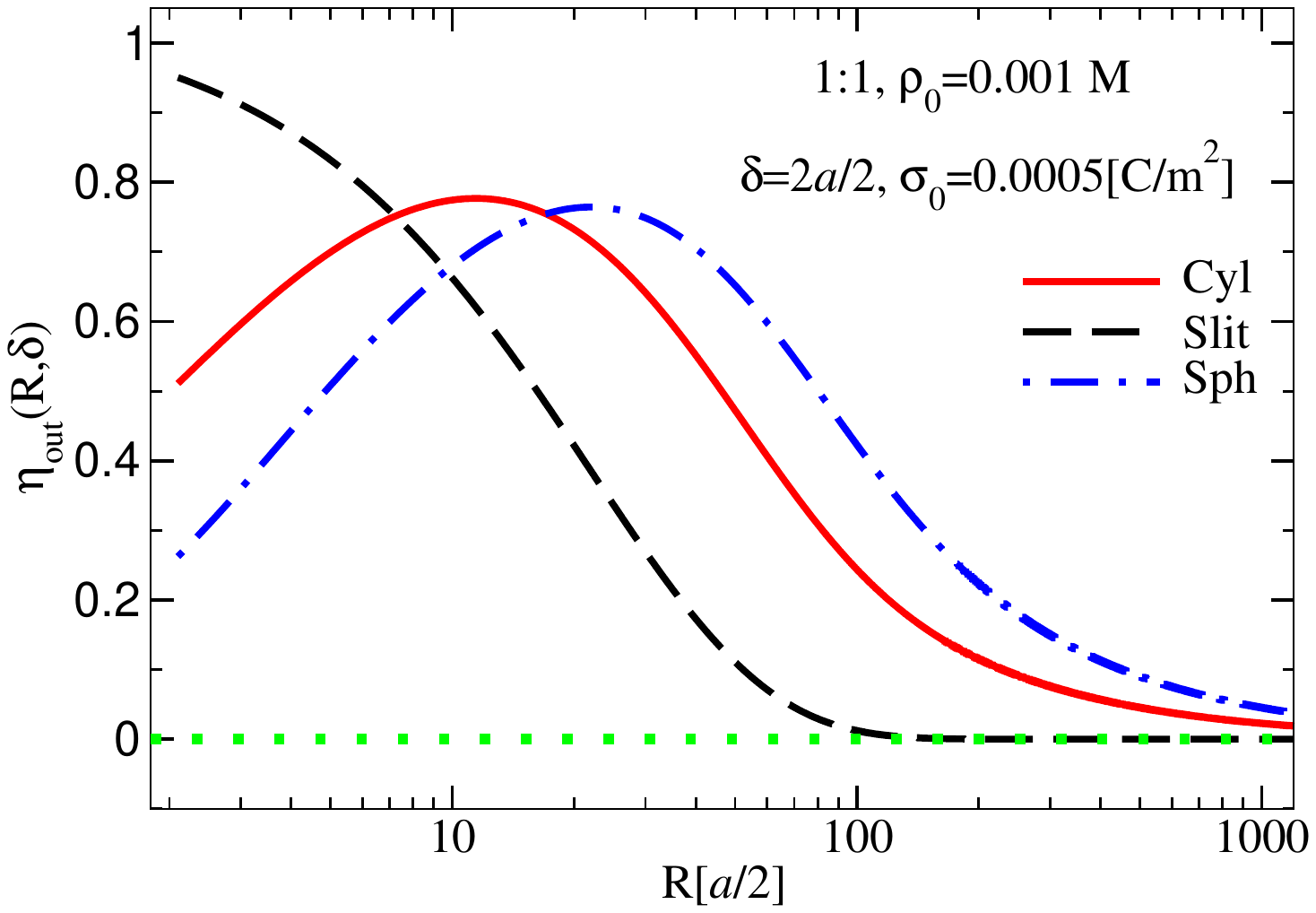}
			\caption{$\eta_{\mathrm{out}}(R,\delta)$}
			\label{fig:EtaOut-d2-s0p0005-rho0p001}
		\end{subfigure}
		
		\vspace{0.3cm}
		
		\begin{subfigure}[b]{0.48\columnwidth}
			\centering
			\includegraphics[width=\linewidth]
			{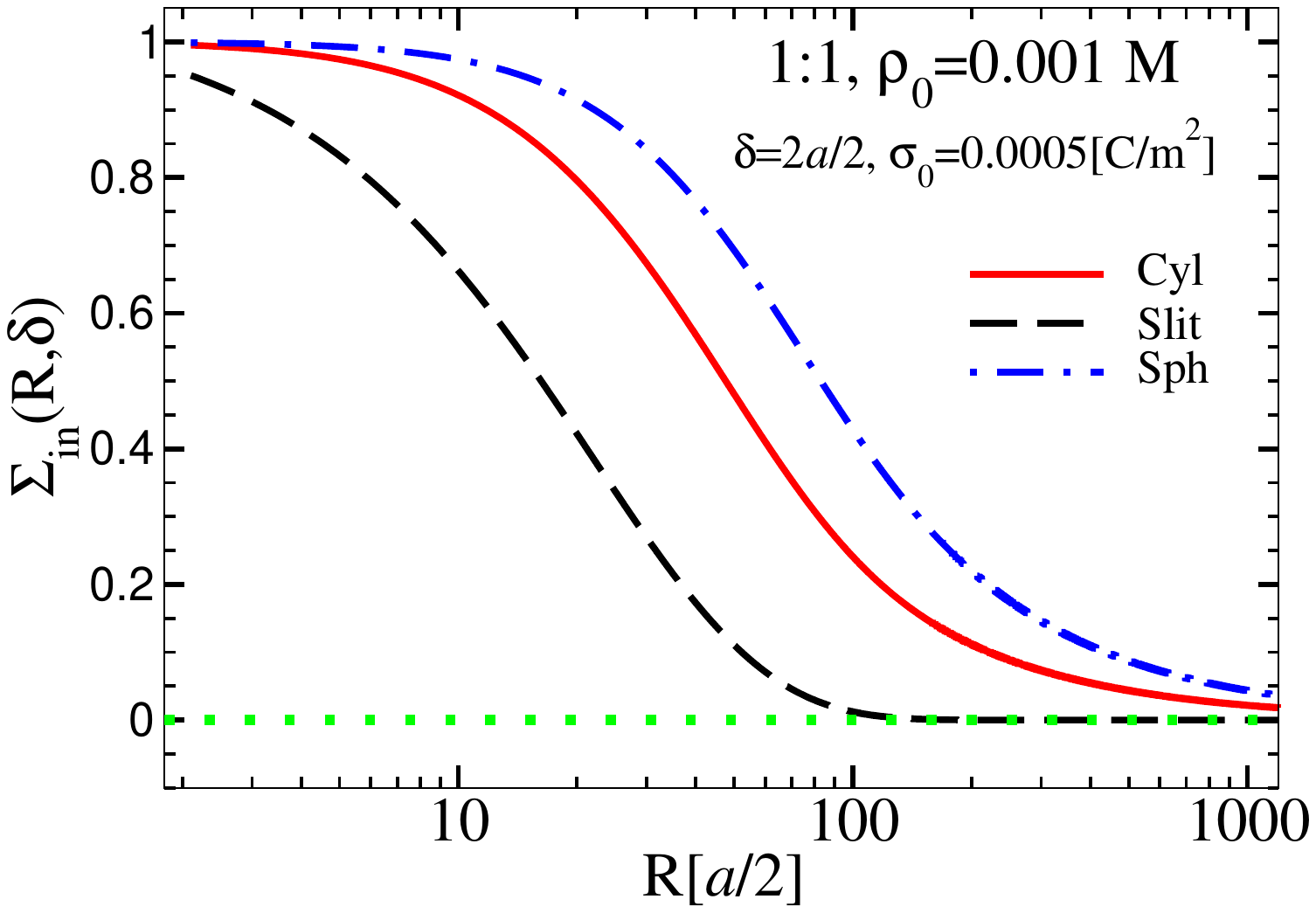}
			\caption{$\Sigma_{\mathrm{in}}(R,\delta)$}
			\label{fig:Delta-SigmaIn-d2-s0p0005-rho0p001}
		\end{subfigure}
		\hfill
		\begin{subfigure}[b]{0.48\columnwidth}
			\centering
			\includegraphics[width=\linewidth]
			{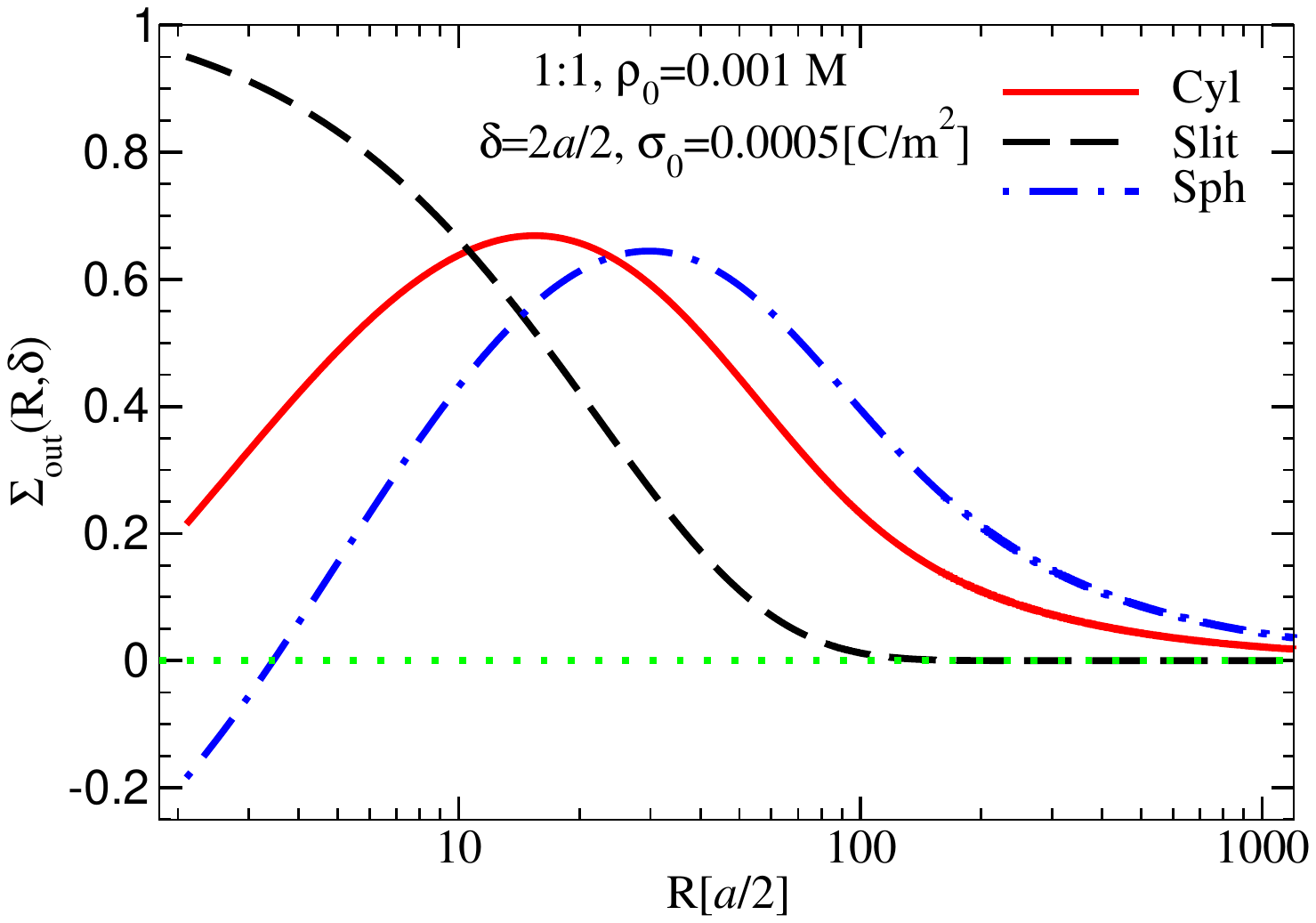}
			\caption{$\Sigma_{\mathrm{out}}(R,\delta)$}
			\label{fig:Delta-SigmaOut-d2-s0p0005-rho0p001}
		\end{subfigure}
		\caption{
			(a) $\eta_{\mathrm{in}}(R,\delta)$,
			(b) $\eta_{\mathrm{out}}(R,\delta)$,
			(c) $\Sigma_{\mathrm{in}}(R,\delta)$, and
			(d) $\Sigma_{\mathrm{out}}(R,\delta)$
			for planar, cylindrical, and spherical hollow nanoparticles with thin walls and very low surface  charge density, i.e. $\delta=2a/2$ and $\sigma_{0}=0.0005 C/m^2$, immersed in a very low electrolyte concentration, $\rho_0=0.001 M$.
		}
		\label{fig:Delta-Sigma-Eta-d2-s0p0005-rho0p001}
		\end{figure}
	

The observed hierarchy originates from the topology of the confining domain. Spherical shells correspond to compact manifolds that enclose a finite volume ($\Omega_{\mathrm{sph}}\simeq S^{2}\times[0,\delta]$), enforcing global electrostatic constraints imposed by Gauss' law. This global enclosure couples the entire charge distribution and produces the strongest violation of local electroneutrality. Cylindrical geometries are only partially compact ($\Omega_{\mathrm{cyl}}\simeq S^{1}\times\mathbb{R}\times[0,\delta]$) and therefore exhibit an intermediate behavior. In contrast, planar slits are non-compact ($\Omega_{\mathrm{slit}}\simeq \mathbb{R}^{2}\times[0,\delta]$), so that the electrostatic response is much less constrained by the global topology. Although $\eta_{\mathrm{T}}(R,\delta)$ decreases monotonically toward its asymptotic value of $0.5$ for all three geometries, the relaxation is substantially faster for planar slits, whereas cylindrical and spherical confinement exhibit a broader crossover and a much more persistent global electroneutrality imbalance. Thus, while VLEC is a consequence of confinement, topology controls its scaling behavior and nonlinear response.

In \cref{fig:etaT-d2-s0p005-rho0p1-z1,fig:etaT-d200-s0p005-rho0p1-z1} we see that increasing $\delta$ increases the amount and range of the VLEC, as a function of $R$, as well as decreasing $\rho_0$ (see \cref{fig:etaT-d2-s0p005-rho0p1-z1,fig:EtaT-d2-s0p005-rho0p001-z1}), while decreasing $\sigma_{\mathrm{0}}$ (see \cref{fig:etaT-d200-s0p005-rho0p1-z1,fig:EtaT-d200-s0p0005-rho0p1-z1}) produces no change in $\eta_{\mathrm{T}}(R,\delta)$. In fact $\eta_{\mathrm{T}}(R,\delta)$ is invariant for any value of $\sigma_{\mathrm{0}}$, as mentioned above, which does not mean that there are no changes in the EDL, as we will show later in this section.

The observed topology-controlled confinement hierarchy therefore does not require ion-size effects and already emerges within the point-ion Poisson--Boltzmann description. These analytical results establish the Poisson--Boltzmann benchmark for topology-controlled violations of local electroneutrality. As discussed in the Introduction, any physically consistent theory of confined electrolytes must recover the Poisson--Boltzmann limit at sufficiently low surface charge densities and dilute electrolyte concentrations. Consequently, the topology-controlled confinement hierarchy identified here provides a natural benchmark for more elaborate electrolyte theories incorporating ion-size and many-body correlations. This expectation is consistent with previous HNC/MSA calculations on confined electrolytes, which exhibit the same qualitative topology-controlled confinement hierarchy despite the explicit inclusion of ion-size correlations~\cite{Aguilar_2007}. Moreover, because the hierarchy originates from the global topology of the confining domain rather than from the local interfacial structure of the electrical double layer, we expect it to remain qualitatively unchanged under more general shell geometries, charge distributions, and electrolyte models, provided that the topology of the confining domain is preserved.

Having established the global electroneutrality imbalance through $\eta_{\mathrm{T}}(R,\delta)$, we now analyze how this imbalance is partitioned between the inner and outer electrolyte and between volumetric and interfacial observables.

 In \cref{fig:Delta-Sigma-Eta-d2-s0p0005-rho0p001}  the reduced induced charge, inside, $\eta_{\mathrm{in}}(R,\delta)$, and outside, $\eta_{\mathrm{out}}(R,\delta)$, and the effective induced charge densities, inside, $\Sigma_{\mathrm{in}}(R,\delta)$, and outside, $\Sigma_{\mathrm{out}}(R,\delta)$, the cavity of cylindrical, planar, and spherical hollow nanoparticles are depicted, as a function of the reduced cavity radius, $R/(a/2)$. In this case, we considered thin shell-wall thickness, $\delta =2(a/2)$, and very low surface charge density on both sides of the nanocavity walls, $\sigma_{\mathrm{0}}=0.0005 [C m^{-2}]$. The nanoparticles are assumed to be  immersed in a $1:1$, $0.001 M$ electrolyte.

 In \cref{fig:EtaIn-d2-s0p0005-rho0p001} we see that while here $\eta_{\mathrm{in}}(R,\delta)$ is a monotonic decreasing function of $R$, in \cref{fig:EtaOut-d2-s0p0005-rho0p001}, outside the cavity, the local charge balance, $\eta_{\mathrm{out}}(R,\delta)$, presents maxima for the cylindrical and spherical topologies . The same qualitative behavior is shown in \cref{fig:Delta-SigmaIn-d2-s0p0005-rho0p001,fig:Delta-SigmaOut-d2-s0p0005-rho0p001}, respectively. Figures \ref{fig:EtaIn-d2-s0p0005-rho0p001} and \ref{fig:EtaOut-d2-s0p0005-rho0p001} are a measure of the charge imbalance inside and outside the cavity of the hollow nanoparticle, with respect to $Q_{\mathrm{0}}(R)$ and $Q_{\mathrm{0}}(R+\delta)$, respectively, and hence a measure of the local VLEC. Conversely, \cref{fig:Delta-SigmaIn-d2-s0p0005-rho0p001,fig:Delta-SigmaOut-d2-s0p0005-rho0p001} also quantify the local VLEC, but are also a measure of the effective electric field at $r=(R-a/2)$, $E_{\scriptscriptstyle{\gamma}}[(R-a/2),\delta)]= \sigma_{\scriptscriptstyle{\gamma}}[(R - a/2),\delta)]/(\varepsilon_{\scriptscriptstyle{0}}\varepsilon)$, and at $r=R_{\mathrm{H}}$, $E_{\scriptscriptstyle{\gamma}}[(R_{\scriptscriptstyle{H}},\delta)]=\sigma_{\scriptscriptstyle{\gamma}}[R_{\scriptscriptstyle{H}},\delta]/(\varepsilon_{\scriptscriptstyle{0}}\varepsilon))$ respectively, where $R_{\scriptscriptstyle{H}}=R +  \delta + a/2$.

As in \cref{fig:Eta2-d2-d200-sig0p005-s0p0005-rho0p1-rho0.001}, for $\eta_{\mathrm{T}}(R,\delta)$, $\eta_{\mathrm{in}}(R,\delta)$, decreases monotonically, for the three nanoparticles' geometries, corresponding to the separate compensation of the fixed charges on the inner and outer shell surfaces by their respective induced charges, and hence, giving a measure of the local VLEC. The closer $\eta_{\mathrm{in}}(R,\delta)$ is to $1$, the larger the VLEC, but $\lim_{R\to\infty}\eta_{\mathrm{in}}(R,\delta)=0$, implying electroneutrality inside the cavity, in this limit. Also the VLEC, of the three topologies exhibits a clear and systematic ordering across them. The non-monotonic behavior of $\eta_{\mathrm{out}}(R,\delta)$ is in response to the VLEC inside the cavity and geometrical effects, strongly dependent also on $\delta$, as we will show later in this article. 

In \cref{fig:Delta-SigmaIn-d2-s0p0005-rho0p001} the electric field, $\sigma_{\mathrm{\gamma}}((R-a/2),a/2)/(\varepsilon_0 \varepsilon)$, at $r=(R-a/2)$ is always negative and lower in magnitude than that on the inner side of the shell's wall, $\sigma_{\mathrm{0}}/(\varepsilon_0 \varepsilon)$, at $r=R$, for $\gamma =0,1,2$. Hence, $\Sigma_{\mathrm{in}}(R,\delta)>0$, for the three geometries, evidencing the charge VLEC also exhibited in \cref{fig:EtaIn-d2-s0p0005-rho0p001}. In \cref{fig:Delta-SigmaOut-d2-s0p0005-rho0p001}, $\Sigma_{\mathrm{out}}(R,\delta)$, for $\gamma =1, 2$ exhibit maxima. Moreover, for $\gamma=0,1$, $\Sigma_{\mathrm{out}}(R,\delta)>0, \forall R>a$. This is nearly the case for $\gamma=2$, except for $R\lesssim3a/2$. A $\Sigma_{\mathrm{out}}(R,\delta)>0,$ implies \textit{confinement overcharging} (CO), as previously reported~\cite{Lozada-Cassou_JML-2025}. This phenomenon is particularly interesting because it is independent of particle-size correlations. In the past, overcharging (also referred to as Surface Charge Amplification (SCA)) has been reported in inhomogeneous macroions solutions~\cite{Jimenez_2004_Feb,Gonzalez-overcharging-cyl-macroions-2022}. However, for macroions, overcharging is a configurational entropy effect. Hence, this CO is a pure confinement and topological phenomena.

As for $\eta_{\mathrm{out}}(R,\delta)$, with respect to $\eta_{\mathrm{in}}(R,\delta)$, in \cref{fig:EtaIn-d2-s0p0005-rho0p001,fig:EtaOut-d2-s0p0005-rho0p001}, $\Sigma_{\mathrm{out}}(R,\delta)$ closely follows that of $\eta_{\mathrm{out}}(R,\delta)$ but depends strongly on the shell-wall thickness, $\delta$, as willl be discussed below. The location of the maxima in $\eta_{\mathrm{out}}(R,\delta)$ depends on the degree of the VLEC and $\delta$, i.e., the lesser and range of the VLEC, because of increasing the salt's concentration and/or valence, i.e., the lower, in intensity and range of $\eta_{\mathrm{in}}(R,\delta)$, as a function of $R$, and the more these maxima shift to smaller cavity sizes.

	\begin{figure}[!htbp]
	\centering
	
	\begin{subfigure}[b]{0.48\columnwidth}
		\centering
		\includegraphics[width=\linewidth]
		{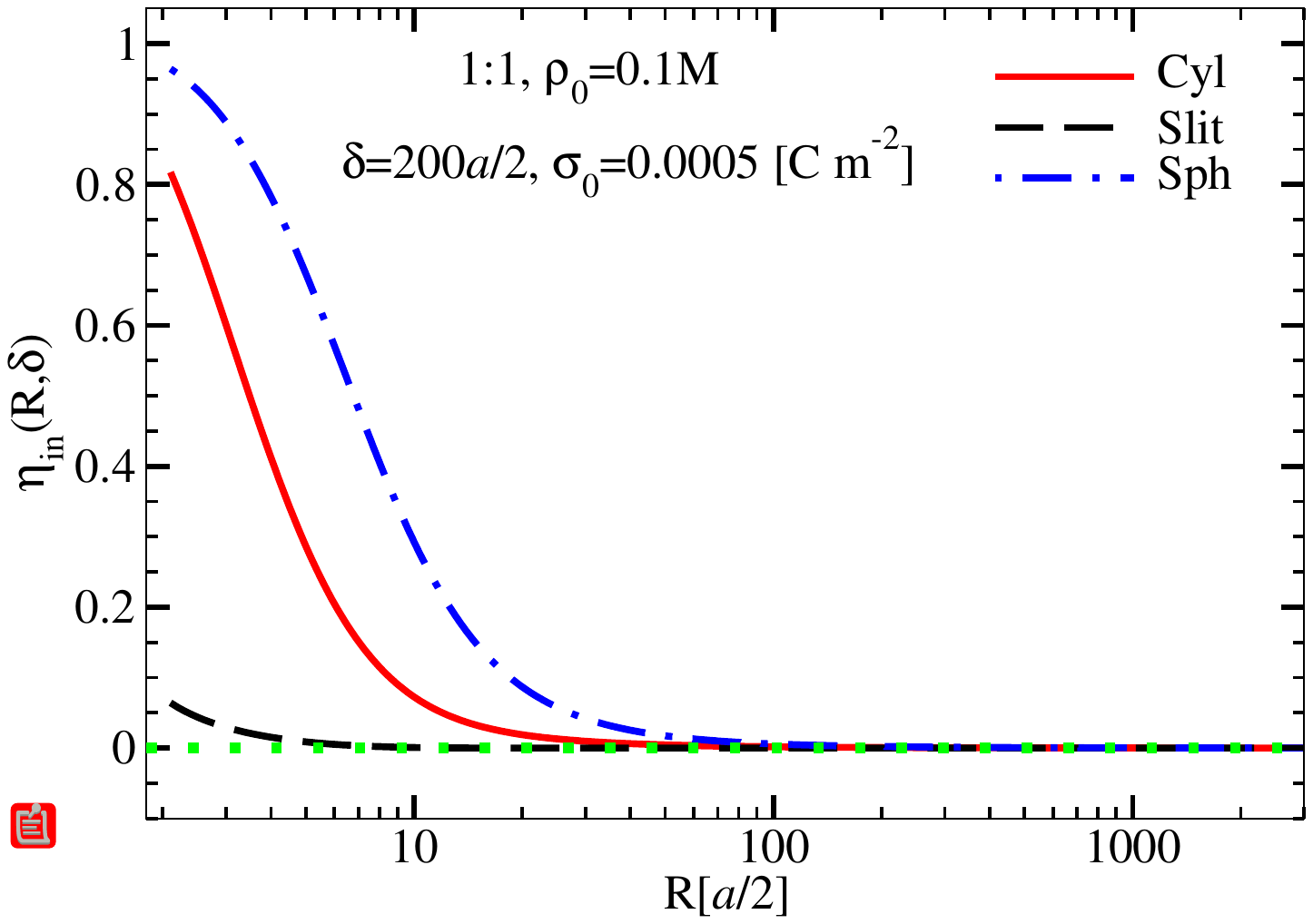}
		\caption{$\eta_{\mathrm{in}}(R,\delta)$}
		\label{fig:EtaIn-d200-s0p0005-rho0p1}
	\end{subfigure}
	\hfill
	\begin{subfigure}[b]{0.48\columnwidth}
		\centering
		\includegraphics[width=\linewidth]
		{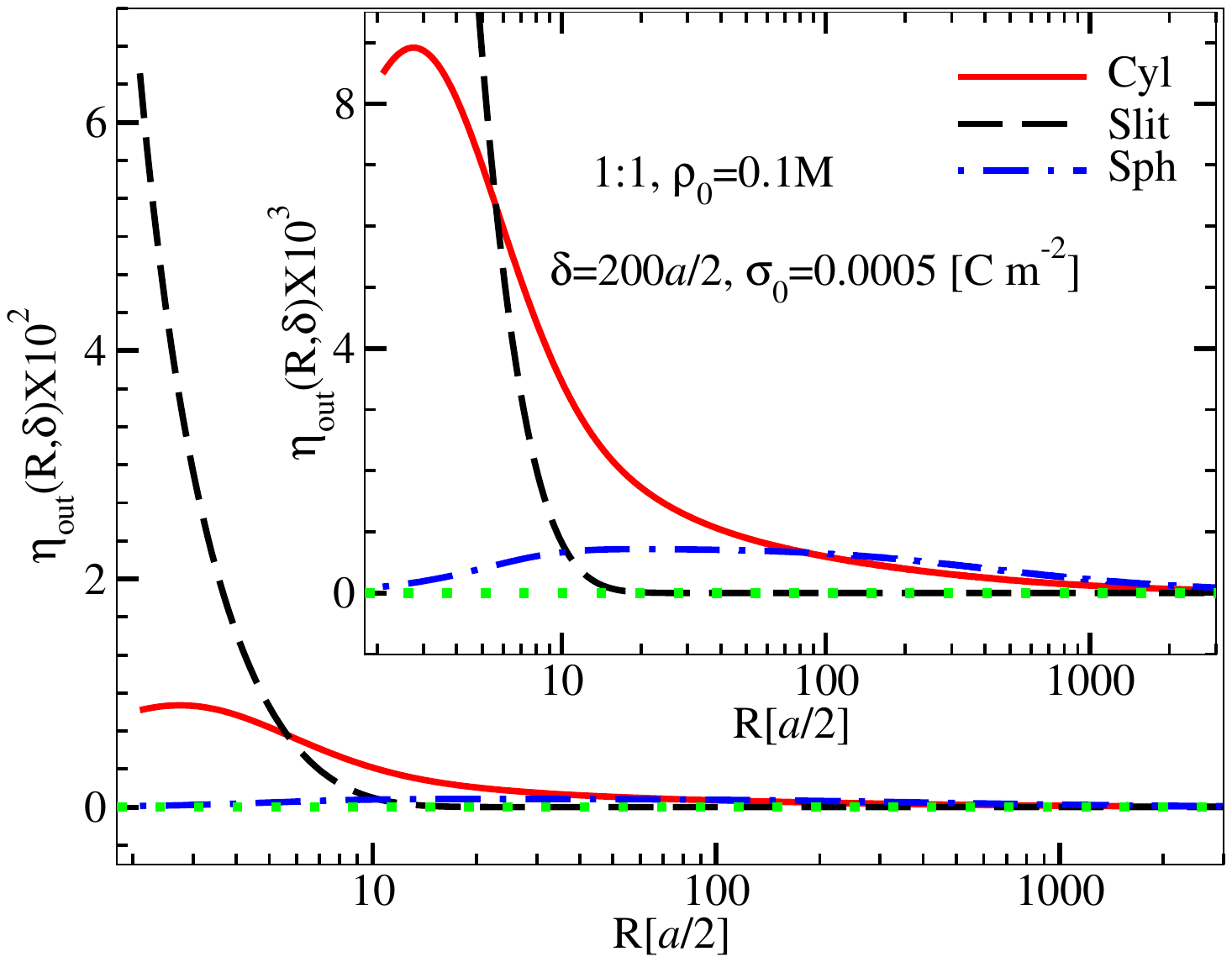}
		\caption{$\eta_{\mathrm{out}}(R,\delta)$}
		\label{fig:EtaOut-d200-s0p0005-rho0p1}
	\end{subfigure}
	
	\vspace{0.3cm}
	
	\begin{subfigure}[b]{0.48\columnwidth}
		\centering
		\includegraphics[width=\linewidth]
		{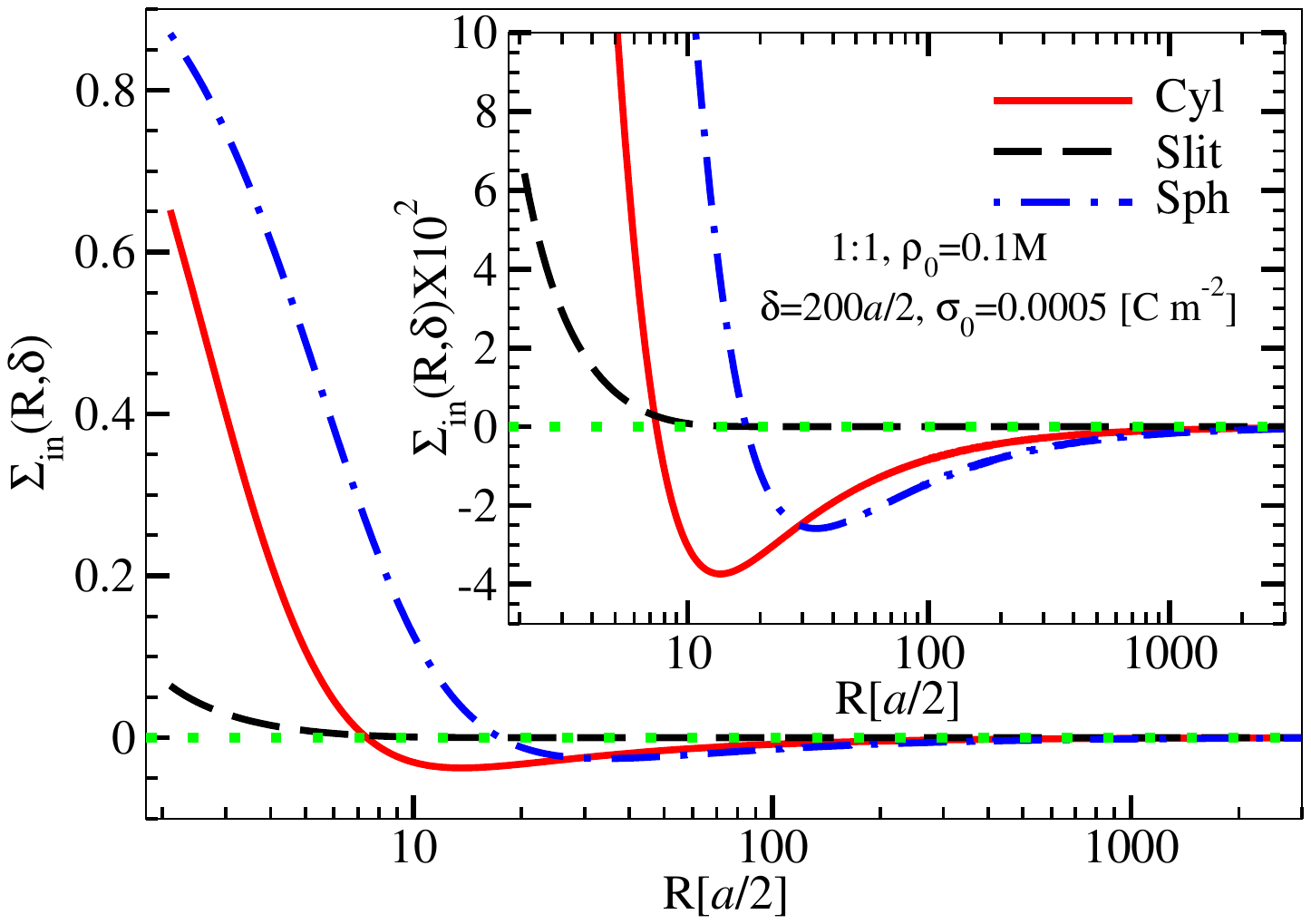}
		\caption{$\Sigma_{\mathrm{in}}(R,\delta)$}
		\label{fig:Delta-Bis2-SigmaIn-d200-s0p0005-rho0p1}
	\end{subfigure}
	\hfill
	\begin{subfigure}[b]{0.48\columnwidth}
		\centering
		\includegraphics[width=\linewidth]
		{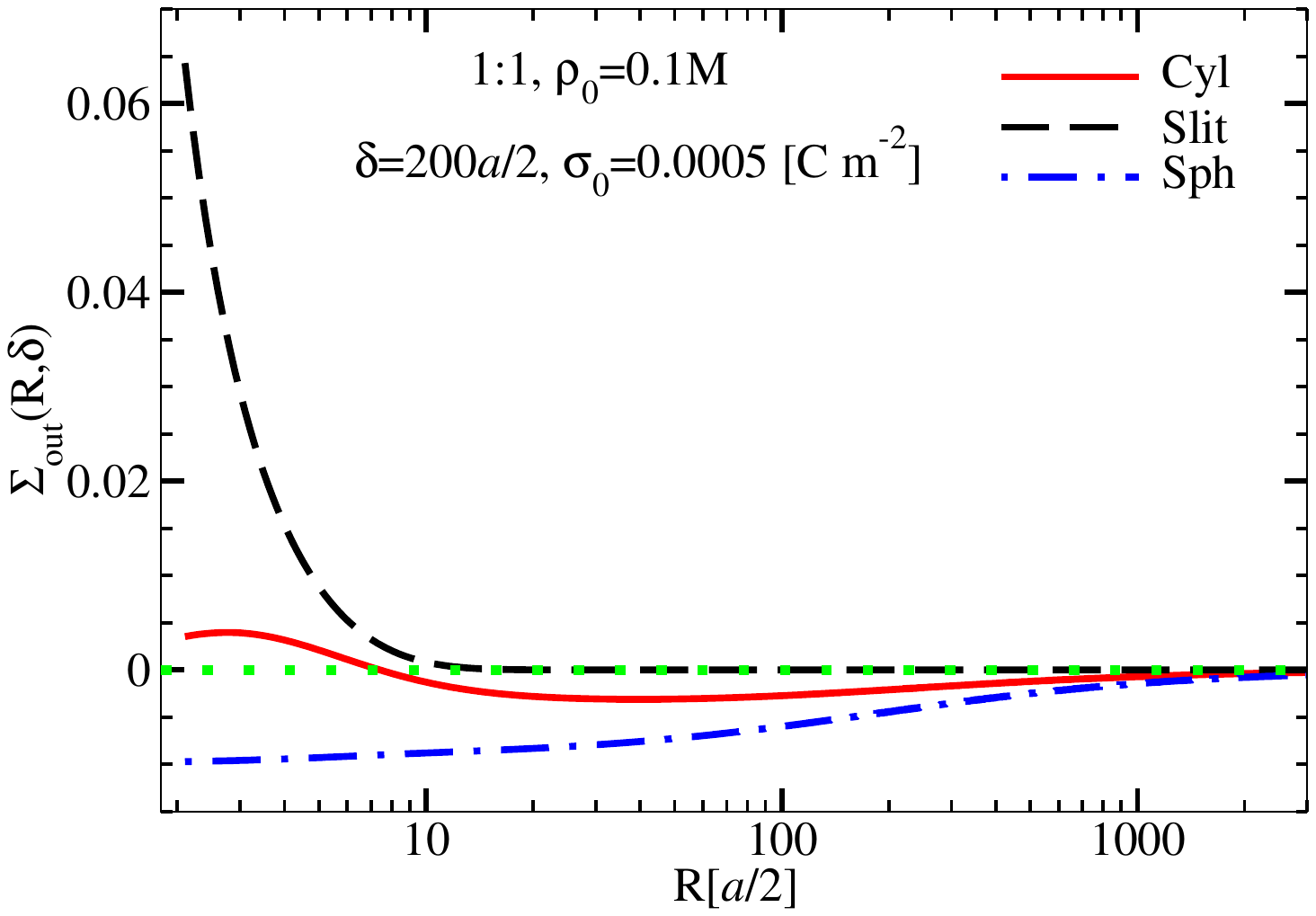}
		\caption{$\Sigma_{\mathrm{out},\delta}(R)$}
		\label{fig:Delta-SigmaOut-d200-s0p0005-rho0p1}
	\end{subfigure}
	\caption{
		(a) $\eta_{\mathrm{in}}(R,\delta)$,
		(b) $\eta_{\mathrm{out}}(R,\delta)$,
		(c) $\Sigma_{\mathrm{in}}(R,\delta)$, and
		(d) $\Sigma_{\mathrm{out}}(R,\delta)$
		for planar, cylindrical, and spherical hollow nanoparticles with thick walls and very low surface  charge density, i.e. $\delta=200a/2$ and $\sigma_{0}=0.0005 C/m^2$, immersed in a low electrolyte concentration, $\rho_0=0.1 M$.}
	\label{fig:DeltaBis-Sigma-EtaBis-d200-s0p0005-rho0p1}
\end{figure}

In \cref{fig:DeltaBis-Sigma-EtaBis-d200-s0p0005-rho0p1}, as in \cref{fig:Delta-Sigma-Eta-d2-s0p0005-rho0p001}, we plot $\eta_{\mathrm{in}}(R,\delta)$, $\eta_{\mathrm{out}}(R,\delta)$, $\Sigma_{\mathrm{in}}(R,\delta)$, and $\Sigma_{\mathrm{out}}(R,\delta)$, as a function of the reduced cavity radius, $R/(a/2)$, for $\gamma=0,1,2$. However, in this case, we considered a thick shell-wall thickness, $\delta =200(a/2)$ and a higher monovalent electrolyte concentration, $\rho_0=0.1 M$. The surface charge density on the wall is also $\sigma_{\mathrm{0}}=0.0005 [C m^{-2}]$. 

In \cref{fig:EtaIn-d200-s0p0005-rho0p1}, $\eta_{\mathrm{in}}(R,\delta)$ preserves the same ordering for the three geometries, as in \cref{fig:EtaIn-d2-s0p0005-rho0p001}, but their range, as a function of $R$, becomes much shorter than those depicted in \cref{fig:EtaIn-d2-s0p0005-rho0p001}, particularly for the slit. This is mainly due to the higher salt concentration, not to the larger value of $\delta$, as can be inferred from \cref{fig:Eta2-d2-d200-sig0p005-s0p0005-rho0p1-rho0.001}.

As,  in \cref{fig:EtaOut-d2-s0p0005-rho0p001}, in \cref{fig:EtaOut-d200-s0p0005-rho0p1} $\eta_{\mathrm{out}}(R,\delta)$ presents maxima, for $\gamma=1,2$, and all exhibit a shorter range. However, let us here point out that while these shorter ranges imply less intensity of the local VLEC, in both cases, i.e., in \cref{fig:EtaOut-d2-s0p0005-rho0p001,fig:EtaOut-d200-s0p0005-rho0p1}, local electroneutrality is achieved only in the limit of $R \rightarrow \infty$. This is also the case of 
$\eta_{\mathrm{in}}(R,\delta)$, shown in \cref{fig:EtaIn-d2-s0p0005-rho0p001,fig:EtaIn-d200-s0p0005-rho0p1}. Consequently, the aftereffects of the VLEC may be observable for large values of $r$, through $E_{\scriptscriptstyle{\gamma}}[r,\delta)]$ and $g_{\scriptscriptstyle{\gamma j}}(r,\delta)$. 

In \cref{fig:Delta-Bis2-SigmaIn-d200-s0p0005-rho0p1} we see that $\Sigma_{\mathrm{in}}(R,\delta)$ becomes negative for large and very large values of $R$, $\gamma=1,2$. As in the case of the shorter range of $\eta_{\mathrm{in}}(R,\delta)$ and $\eta_{\mathrm{out}}(R,\delta)$, shown in \cref{fig:EtaIn-d2-s0p0005-rho0p001,fig:EtaOut-d2-s0p0005-rho0p001}, with respect to those in \cref{fig:EtaIn-d200-s0p0005-rho0p1,fig:EtaOut-d200-s0p0005-rho0p1}, the emerging of negative values of $\Sigma_{\mathrm{in}}(R,\delta)$ seen in  \cref{fig:Delta-Bis2-SigmaIn-d200-s0p0005-rho0p1} are due mainly to the larger value of $\delta$.

Negative values of $\Sigma_{\mathrm{in}}(R,\delta)$ imply more than just a VLEC, but that the negative induced charge density inside the cavities of the cylindrical and spherical hollow nanoparticles reverses the charge on inner surface of the shell's walls, i.e., \textit{inside charge reversal}.  And since no configurational entropy is present, this phenomenon, as in the case of CO and CCR. is resultant from the confinement topology and the charge correlation of the electrolyte, across the confining walls of the cylindrical and spherical hollow nanoparticles. 

To the best of our knowledge this inside charge reversal has not been reported in the literature, so hereinafter let us refer to this new phenomenon as \textit{inside confinement charge reversal} (ICCR). In contrast, in \cref{fig:Delta-SigmaOut-d200-s0p0005-rho0p1}, $\Sigma_{\mathrm{out}}(R,\delta)$ shows CO for the slit and cylindrical nanoshells, for small values of $R$.

In \cref{fig:EtaT-in-out-Sigma-in-out-d200-s0p005-rho0p2-z2}, we plot $\eta_{\mathrm{T}}(R,\delta)$, $\eta_{\mathrm{in}}(R,\delta)$, $\eta_{\mathrm{out}}(R,\delta)$, $\Sigma_{\mathrm{in}}(R,\delta)$, and $\Sigma_{\mathrm{out}}(R,\delta)$, as a function of the reduced cavity radius, $R/(a/2)$, for cylindrical, spherical and planar nanocavities, as in \cref{fig:DeltaBis-Sigma-EtaBis-d200-s0p0005-rho0p1}, but now for a $2:2$, $0.2 M$ electrolyte, and a higher surface charge density, $\sigma_{\mathrm{0}}=0.005 C/m^2$, while their shell-wall thickness remain the same, i.e., $\delta=200a/2$.

\begin{figure}[!htbp]
	\centering
	\begin{subfigure}[b]{0.48\columnwidth}
		\centering
		\includegraphics[width=\linewidth]
		{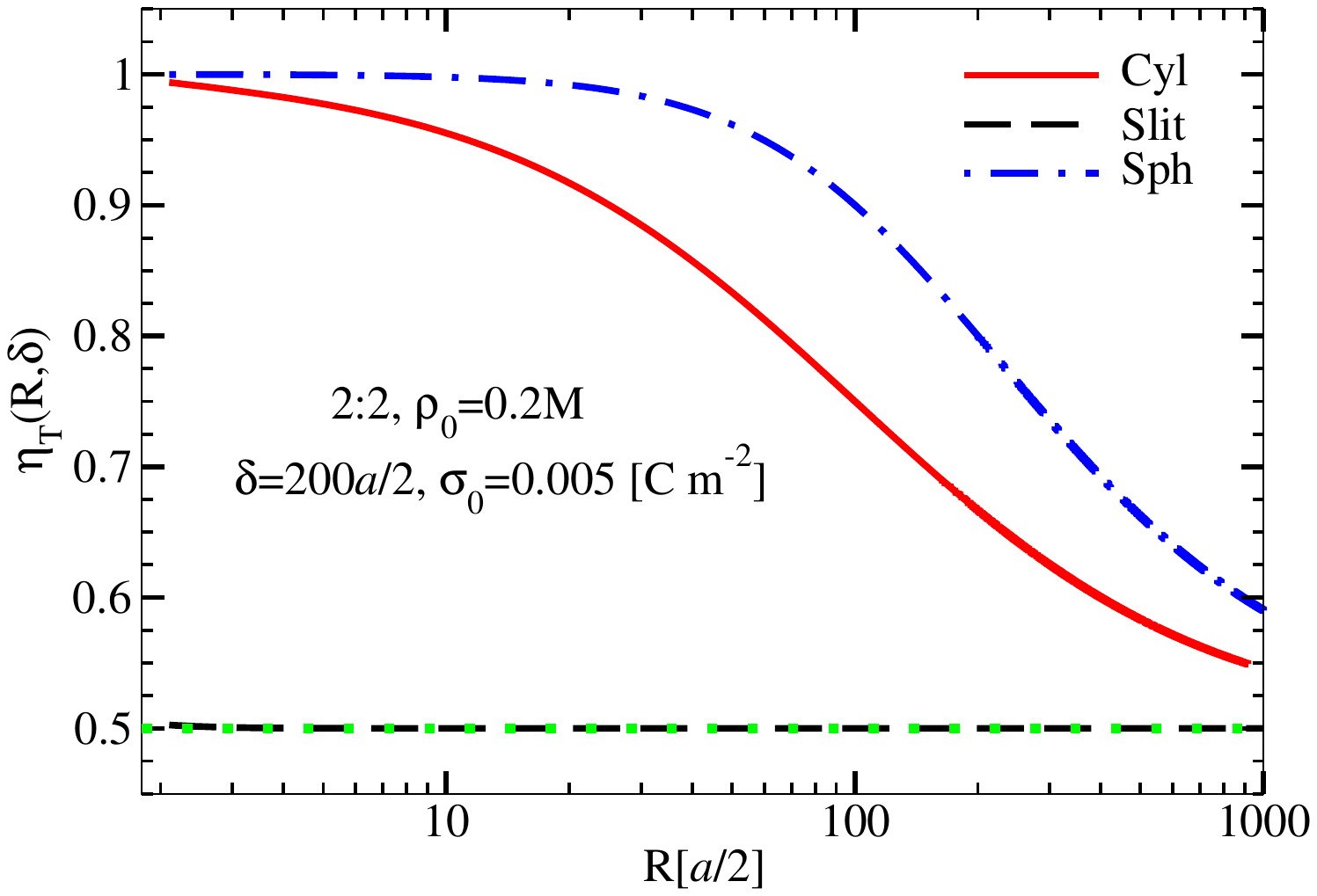}
		\caption{$\eta_{\mathrm{T}}(R,\delta)$}
		\label{fig:Eta2-sigmaHo-0p005_d100_rho0p2_z2-LongR}
	\end{subfigure}
	\hfill
	
	\begin{subfigure}[b]{0.48\columnwidth}
		\centering
		\includegraphics[width=\linewidth]
		{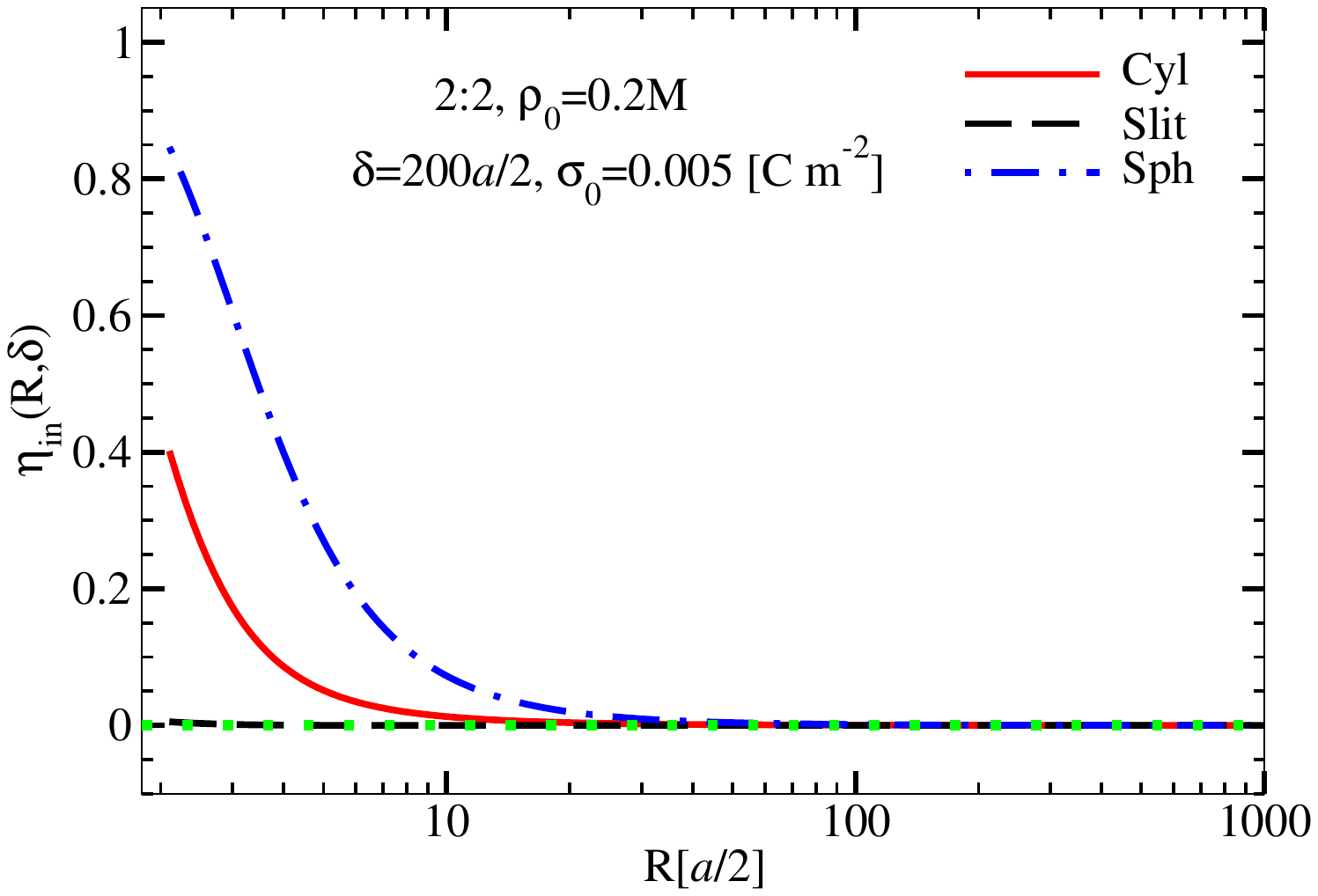}
		\caption{$\eta_{\mathrm{in}}(R,\delta)$}
		\label{fig:EtaIn-d200-s0p005-rho0p2-z2}
	\end{subfigure}
	\hfill
	\begin{subfigure}[b]{0.48\columnwidth}
		\centering
		\includegraphics[width=\linewidth]
		{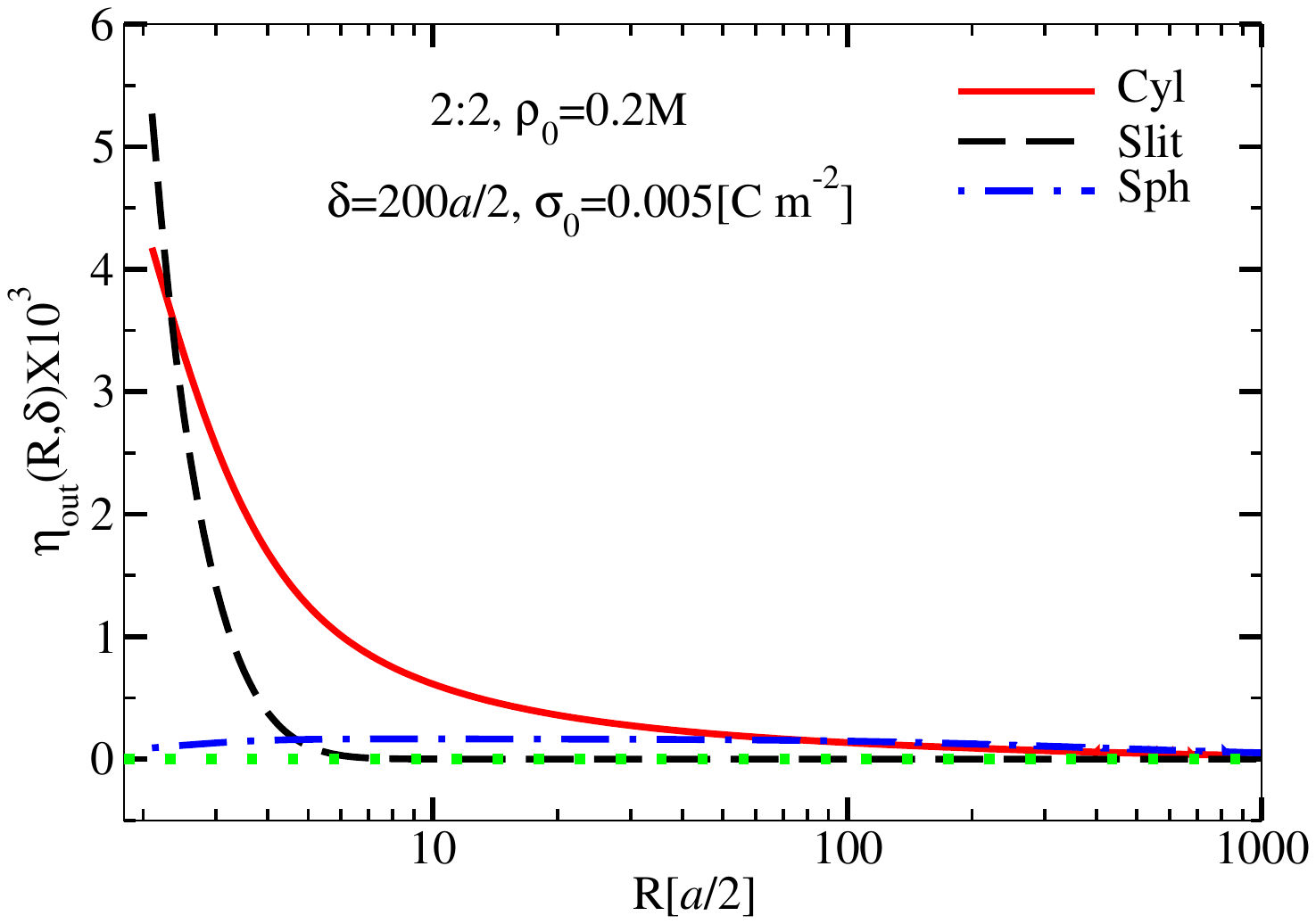}
		\caption{$\eta_{\mathrm{out}}(R,\delta)$}
		\label{fig:EtaOut-d200-s0p005-rho0p2-z2}
		\label{fig:}
	\end{subfigure}
	
	\vspace{0.3cm}
	
	\begin{subfigure}[b]{0.48\columnwidth}
		\centering
		\includegraphics[width=\linewidth]
		{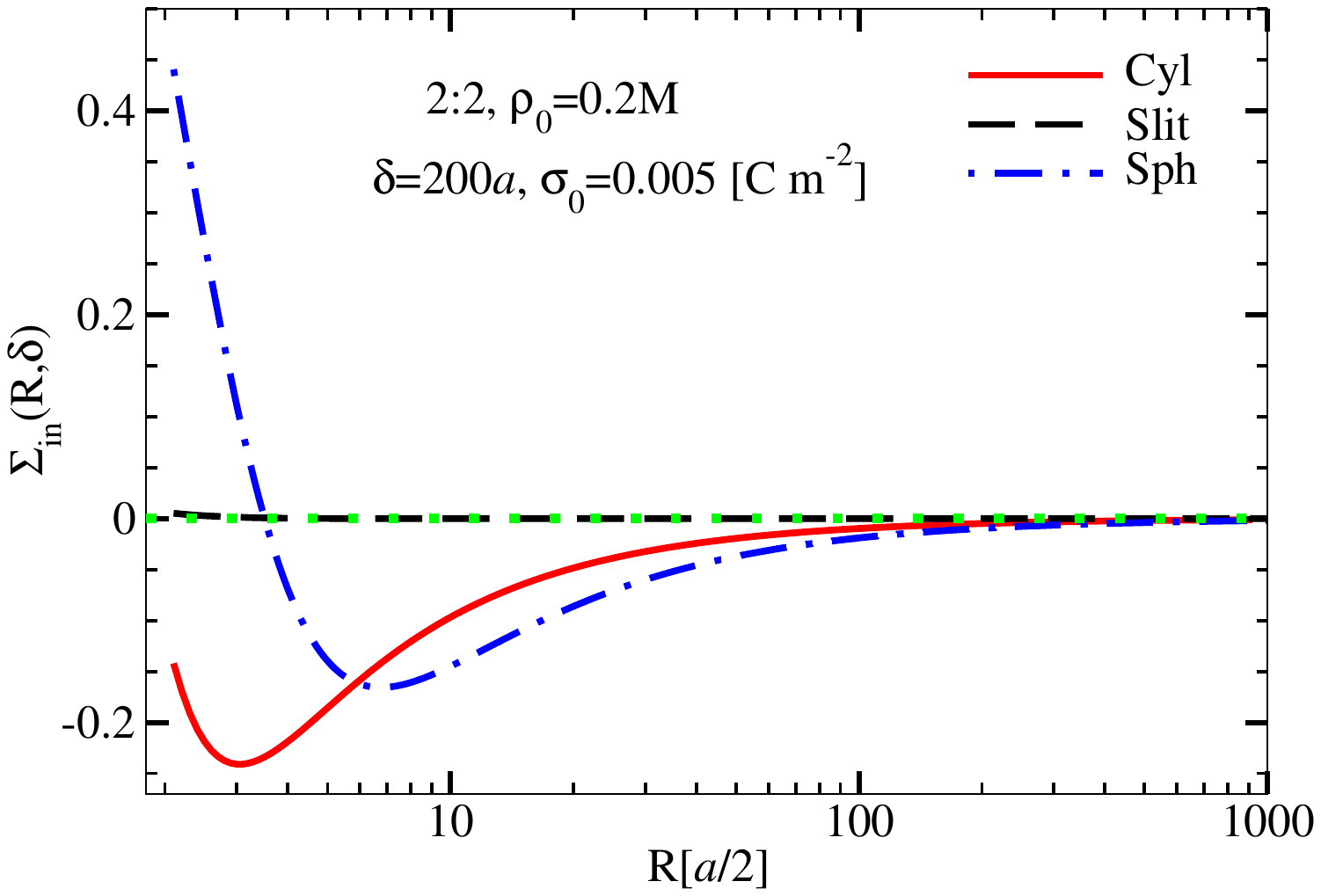}
		\caption{$\Sigma_{\mathrm{in}}(R,\delta)$}
		\label{fig:SigmaIn-d200-s0p005-rho0p2-z2}
	\end{subfigure}
	\hfill
	\begin{subfigure}[b]{0.48\columnwidth}
		\centering
		\includegraphics[width=\linewidth]
		{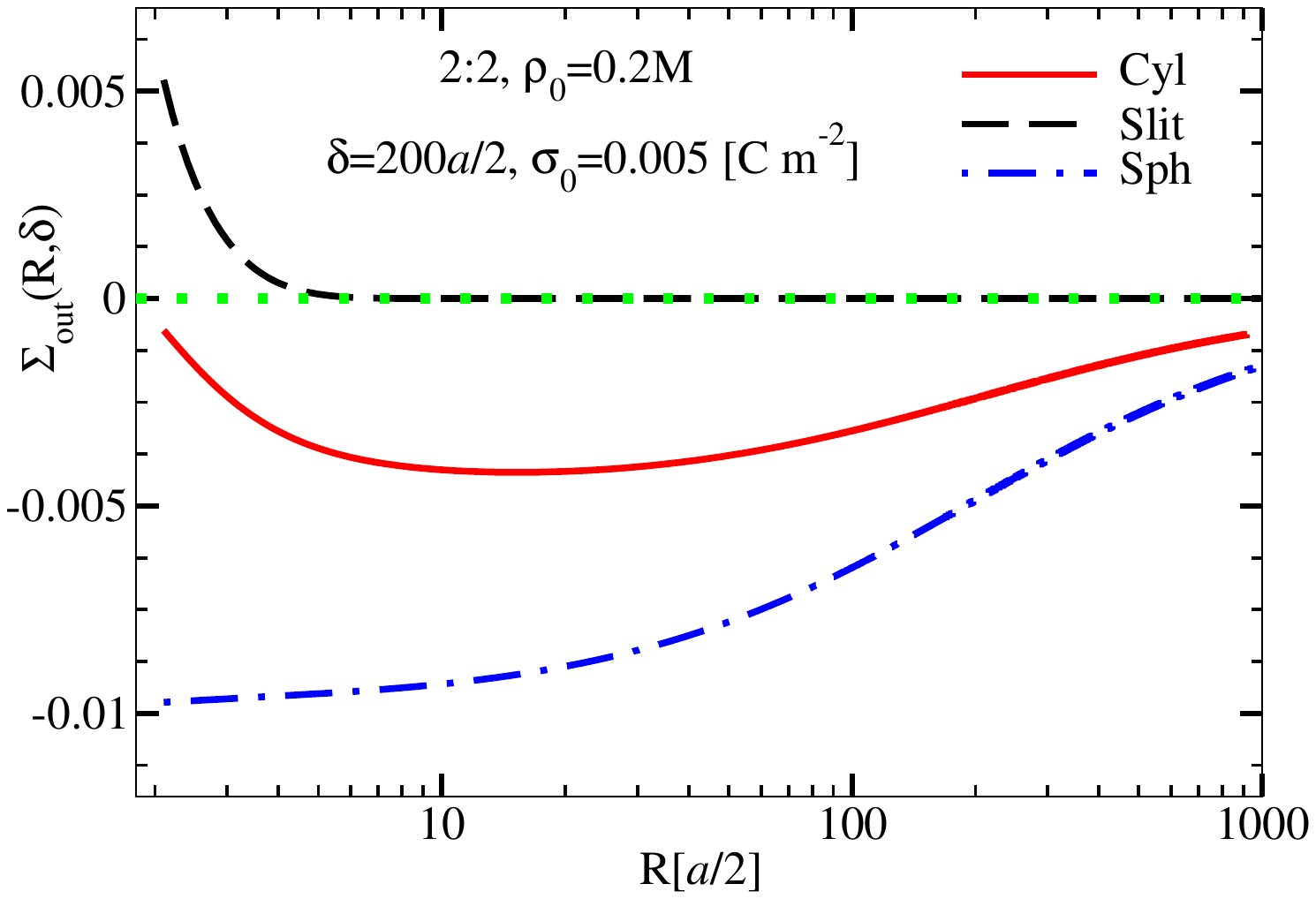}
		\caption{$\Sigma_{\mathrm{out}}(R)$}
		\label{fig:SigmaOut-d200-s0p005-rho0p2-z2}
	\end{subfigure}
	\caption{
		(a) $\eta_{\mathrm{T}}(R,\delta)$,
		(b) $\eta_{\mathrm{in}}(R,\delta)$,
		(c) $\eta_{\mathrm{out}}(R,\delta)$,
		(d) $\sigma_{\mathrm{in}}(R,\delta)$, and
		(e) $\sigma_{\mathrm{out}}(R,\delta)$
		for planar, cylindrical, and spherical hollow nanoparticles with thick walls and higher surface charge density, immersed in a $2:2$, $0.2M$ electrolyte.}
	\label{fig:EtaT-in-out-Sigma-in-out-d200-s0p005-rho0p2-z2}
\end{figure}

A higher electrolyte valence and concentration implies a shorter Debye Length, $\lambda_{\scriptscriptstyle{D}}$ (see \cref{Ec.kappa}); notwithstanding, counterintuitively, the curves of $\eta_{\mathrm{T}}(R,\delta)$, as a function of $R$, is, within undetectable numerical error, identical to those in \cref{fig:etaT-d200-s0p005-rho0p1-z1,fig:EtaT-d200-s0p0005-rho0p1-z1}, for $\gamma=1,2$, thence $\eta_{\mathrm{T}}(R,\delta)$ seems to be invariant on electrolyte valence, as with surface charge density. However, for the slit, $\gamma=0$, $\eta_{\mathrm{T}}(R,\delta)$ becomes much less intense and of shorter range. Nonetheless, their ordering is insensitive also for a 2:2 electrolytes (see \cref{Eq:VLEC-ordering}) and, hence, the topology-controlled confinement hierarchy is preserved, as in \cref{fig:Eta2-d2-d200-sig0p005-s0p0005-rho0p1-rho0.001} for $1:1$ electrolytes. Therefore, the ordering sphere > cylinder > slit is a robust feature of confined electrolytes, as pointed out before. In \cref{fig:EtaT-in-out-Sigma-in-out-d200-s0p005-rho0p2-z2} the range of in $R/(a/2)$ is shorter because of no numerical  convergence of the modified Bessel functions in the electric field equations, for the cylindrical hollow nanoparticle, with these more demanding parameters conditions. 

On the other hand,  a shorter $\lambda_{\scriptscriptstyle{D}}$ do imply a less intense $\eta_{\mathrm{in}}(R,\delta)$ and $\eta_{\mathrm{out}}(R,\delta)$, as seen in  \cref{fig:EtaIn-d200-s0p005-rho0p2-z2,fig:EtaOut-d200-s0p005-rho0p2-z2}, in comparison with \cref{fig:EtaIn-d200-s0p0005-rho0p1,fig:EtaOut-d200-s0p0005-rho0p1}, as expected. This is also the case of $\Sigma_{\mathrm{in}}(R,\delta)$ and $\Sigma_{\mathrm{out}}(R,\delta)$ in \cref{fig:SigmaIn-d200-s0p005-rho0p2-z2,fig:SigmaOut-d200-s0p005-rho0p2-z2}, when compared with \cref{fig:Delta-Bis2-SigmaIn-d200-s0p0005-rho0p1,fig:Delta-SigmaOut-d200-s0p0005-rho0p1}, considering that these functions are invariant to changes in surface charge densities, as discussed above in relation to \cref{,fig:EtaT-d200-s0p0005-rho0p1-z1,fig:etaT-d200-s0p005-rho0p1-z1}. We have calculated these functions with the same parameters, including salt concentrations, depicted in \cref{fig:EtaT-in-out-Sigma-in-out-d200-s0p005-rho0p2-z2}, but for a $1:1$ salt (not shown), and in all cases and, in effect, a shorter $\lambda_{\scriptscriptstyle{D}}$ implies a less intensive charge and surface charge density balance, even though they both become equal to $0.5$, in $\eta_{\mathrm{T}}(R,\delta)$, and equal to $0$, for $\eta_{\mathrm{in}}(R,\delta)$, $\eta_{\mathrm{out}}(R,\delta)$, $\Sigma_{\mathrm{in}}(R,\delta)$, and $\Sigma_{\mathrm{out}}(R,\delta)$, in the $\lim_{R\rightarrow\infty}=0$, as shown in appendix \ref{appendix-B}. Less intensive, shorter range of $\eta_{\mathrm{in}}(R,\delta)$ and $\eta_{\mathrm{out}}(R,\delta)$, imply a lower VLEC.

A shorted $\lambda_{\scriptscriptstyle{D}}$ also implies a more compact EDL, hence, for this divalent, more concentrated salt the ICCR effect shown in \cref{fig:SigmaIn-d200-s0p005-rho0p2-z2} for $\gamma=1,2$ is much more pronounced than that for the monovalent electrolyte in \cref{fig:Delta-Bis2-SigmaIn-d200-s0p0005-rho0p1}. In the case of the slit, no ICCR is present, reaching electroneutrality at $R\simeq5a/2$. On the contrary, in \cref{fig:SigmaOut-d200-s0p005-rho0p2-z2}, $\Sigma_{\mathrm{out}}(R,\delta)\geq0$ for $R\lessapprox 12a/2$ and $\gamma=0$, thence showing some CO, whereas no CO is 
is present for the cylindrical and spherical geometries.

The persistence of these effects, i.e. maxima in $\eta_{\mathrm{out}}(R,\delta)$, CO, CCR and ICCR, until the singular $\lim_{R \to \infty}$, suggests that topology and inside-outside charge correlations become effectively entangled under confinement.

 \section{Discussion}\label{discussion}
 
 Phenomena such as overcharging and charge reversal are attributed to configurational correlations associated with finite ion size and are therefore generally assumed to be absent in point-ion models. In contrast, the present results show that confinement-induced overcharging and charge reversal can emerge even for point ions as a consequence of global electrostatic constraints imposed by confinement. The self-consistent solution of the Poisson--Boltzmann equation couples the entire charge distribution through long-range Coulomb interactions, and this coupling becomes structurally constrained by the topology of the confining domain. As a consequence, nontrivial charge organization can arise within a point-ion theory, reflecting global electrostatic constraints rather than the local correlation effects traditionally invoked to explain these phenomena~\cite{Attard_1996,Vlachy1989,Kjellander-charge-inversion-1998,Netz-Orland-Beyond-PB-2000,Deserno-2001,Grosse-2002,Jimenez_2004_Feb,Green-electroneutrality-JCP-2021,Henderson2012,Gonzalez-overcharging-cyl-macroions-2022,Lozada-Cassou_JML-2025}. These constraints originate from the integral form of Gauss' law applied to domains with different topological properties, and the fact that electrolyte and the hollow nanoparticles form a single statistical-mechanical system.

Topology-induced confinement systematically violates local electroneutrality, and these violations manifest differently when measured through integrated charge, $\eta_{\mathrm{T}}(R,\delta)$, or through local charge density or electric field balances, i.e., $\eta_{\mathrm{in}}(R,\delta)$ and $\eta_{\mathrm{out}}(R,\delta)$ through local charge balance, while $\Sigma_{\mathrm{in}}(R,\delta)$ and $\Sigma_{\mathrm{out}}(R,\delta)$ through local electric field balance, since $\sigma_{\mathrm{in}}(R,\delta)$ and $\sigma_{\mathrm{out}}(R,\delta)$ are a direct measure of the local electric fields $E_{\scriptscriptstyle{\gamma}}[(R-a/2),\delta)]= \sigma_{\scriptscriptstyle{\gamma}}[(R - a/2),\delta)]/(\varepsilon_{\scriptscriptstyle{0}}\varepsilon)$, and $E_{\scriptscriptstyle{\gamma}}[(R_{\scriptscriptstyle{H}},\delta)]=\sigma_{\scriptscriptstyle{\gamma}}[R_{\scriptscriptstyle{H}},\delta]/(\varepsilon_{\scriptscriptstyle{0}}\varepsilon))$

 These results show that local electric fields do not scale identically to integrated charges, especially in cylindrical and spherical geometries. A key conceptual point is: While integrated charges quantify the violation of local charge electroneutrality, the corresponding surface charge densities quantifies the violation of local electric field electroneutrality. These two measures are not equivalent, since area factors introduce geometry-dependent scaling. Thus, spherical and cylindrical confinement may display similar integrated charge compensation but markedly different local field strengths.

\section{Conclusions}\label{conclusions}
In summary, we have shown that the finite-size violation of local electroneutrality in confined electrolytes is governed by the topology of the confining domain. The resulting hierarchy of deviations across spherical, cylindrical, and planar geometries demonstrates that charge redistribution under confinement is controlled by global electrostatic constraints, rather than by local geometric details or particle models. These findings establish topology as a fundamental organizing principle in the electrostatics of confined systems. The same topological hierarchy governs not only the integrated charge imbalance but also the local charge and electric-field balances.

More broadly, the existence of VLEC reflects a global electrostatic constraint that determines the electrostatic structure of both the confined region and the surrounding electrolyte. This constraint influences thermodynamic observables such as osmotic pressure, as well as adsorption phenomena in biological systems (e.g., cells and vesicles) and collective behavior in colloidal systems, highlighting the broad relevance of VLEC across soft condensed matter and biological contexts.

Although this study is limited to symmetric electrolytes within the linear Poisson–Boltzmann approximation, our methodology can be extended to asymmetric systems in ion size and valence, as well as constant charge, and asymmetrically charged hollow nanoparticles' wall charge distributions, which would be relevant for the cylindrical and spherical geometries. The analysis of hollow nanoparticle capacitance may have implications not only for energy storage device design, but also for understanding micelles, nanopores, and confined charged systems in biology, medicine, drug delivery, and the chemical industry.

Finally, our results suggest a broader implication. The magnitude of the VLEC is governed by the topology, rather than the particular geometry, of the confining domain, as illustrated by the topological classes
$\Omega_{\mathrm{slit}}\simeq\mathbb{R}^{2}\times[0,\delta]$,
$\Omega_{\mathrm{cyl}}\simeq S^{1}\times\mathbb{R}\times[0,\delta]$,
and
$\Omega_{\mathrm{sph}}\simeq S^{2}\times[0,\delta]$.
Consequently, systems with markedly different morphologies may be governed by the same global electrostatic constraints, provided they belong to the same topological class. This viewpoint may therefore prove useful in biological systems exhibiting large morphological variability, where cells with very different shapes—for example, pleomorphic, spindle-shaped, or squamous cancer cells—may belong to the same topological class. More generally, this perspective extends naturally to confined electrolytes and soft condensed matter systems in which topology, rather than geometric detail, controls the electrostatic response.

From a theoretical point of view, this study demonstrates that changing the topology modifies the global electrostatic constraints without altering the local Poisson--Boltzmann equations. This shows that topology enters electrolyte theory not through the local differential equations themselves, but through the global constraints imposed on their self-consistent solutions. Since the electrolyte and the hollow nanoparticle constitute a single statistical-mechanical system, these constraints organize the electrostatic response throughout the entire confined system.

\textit{Acknowledgements.—}
The author gratefully acknowledges support from DGAPA-UNAM through PAPIIT Project No.~ IN108023.

\appendix

\section{Electrical field}
\label{appendix-A}

Analytical solutions of \cref{Laplace-equation,eq:LPB} for the three hollow nanoparticle geometries have been previously derived~\cite{Adrian-JML-2023}. Here, we present a reformulation of the relevant expressions for the electrostatic potential \( \psi(r) \) and the electric field \( E(r) \), specific to each geometry. We define \( \sigma_{\scriptscriptstyle{Hi}} = \sigma(R - a/2) \) and \( \sigma_{\scriptscriptstyle{Ho}} = \sigma(R + \delta + a/2) \) as the \textit{effective induced surface charge densities} at the inner and outer electrolyte interfaces, respectively.

These expressions are exact within the LPB framework. The discontinuities in the electric field at \( r = R \) and \( r = R + d \delta\) arise from Maxwell’s boundary conditions at the interfaces between the electrolyte and the charged shell wall.

\subsubsection{Slit-Shell}\label{Slit-shell}

The LPB solution for a symmetric electrolyte in a slit-shell geometry (i.e., two parallel plates separated by \( 2R \)) with a Stern layer correction is given by~\cite{Adrian-JML-2023}:

\paragraph{Mean Electrostatic Potential \(\psi(r)\):}

\begin{equation}
	A_p=
	\frac{2\sigma_0-\sigma_{\scriptscriptstyle Ho}}
	{\varepsilon_{\scriptscriptstyle 0}\varepsilon},
	\qquad
	B_p=
	\frac{\sigma_{\scriptscriptstyle Ho}}
	{\varepsilon_{\scriptscriptstyle 0}\varepsilon},
	\qquad
	C_p=
	\frac{\sigma_{\scriptscriptstyle Ho}-\sigma_0}
	{\varepsilon_{\scriptscriptstyle 0}\varepsilon}
\end{equation}
\begin{equation}
	\psi(r=|x|) =
	\begin{cases}
		\displaystyle
		\frac{A_p}{\kappa}
		\frac{\cosh(\kappa r)}
		{\sinh[\kappa(R-a/2)]},
		&
		0\leq r\leq R-a/2,
		\\[8pt]
		
		\displaystyle
		\psi_{\scriptscriptstyle H}
		+
		A_p\big[r-(R-a/2)\big],
		&
		R-a/2\leq r\leq R,
		\\[8pt]
		
		\displaystyle
		\varphi_0-C_p(r-R- \delta),
		&
		R\leq r\leq R+ \delta,
		\\[8pt]
		
		\displaystyle
		\frac{B_p}{\kappa}
		\big[1-\kappa(r-R_{\scriptscriptstyle H})\big],
		&
		R+ \delta\leq r\leq R_{\scriptscriptstyle H},
		\\[8pt]
		
		\displaystyle
		\frac{B_p}{\kappa}
		e^{-\kappa(r-R_{\scriptscriptstyle H})},
		&
		R_{\scriptscriptstyle H}\leq r .
	\end{cases}
	\label{Plates-MEP-r}
\end{equation}


\paragraph{Electric Field \(E(r)\):}

\begin{equation}
	E(r) =
	\begin{cases}
		\displaystyle \frac{\sigma_{\scriptscriptstyle{Ho}} - 2\sigma_0}{\varepsilon_{\scriptscriptstyle{0}}\varepsilon} \frac{\sinh[\kappa r]}{\sinh[\kappa(R - a/2)]}, & 0 \leq r \leq R - \frac{a}{2} \\[8pt]
		\displaystyle \frac{\sigma_{\scriptscriptstyle{Ho}} - 2\sigma_0}{\varepsilon_{\scriptscriptstyle{0}}\varepsilon}, & R - \frac{a}{2} \leq r \leq R \\[8pt]
		\displaystyle \frac{\sigma_{\scriptscriptstyle{Ho}} - \sigma_0}{\varepsilon_{\scriptscriptstyle{0}}\varepsilon}, & R < r < R +  \delta \\[8pt]
		\displaystyle \frac{\sigma_{\scriptscriptstyle{Ho}}}{\varepsilon_{\scriptscriptstyle{0}}\varepsilon}, & R +  \delta \leq r \leq R_{\scriptscriptstyle{H}} \\[8pt]
		\displaystyle \frac{\sigma_{\scriptscriptstyle{Ho}}}{\varepsilon_{\scriptscriptstyle{0}}\varepsilon} e^{-\kappa(r - R_{\scriptscriptstyle{H}})}, & R_{\scriptscriptstyle{H}} \leq r
	\end{cases}
	\label{Plates-E(r)}
\end{equation}
Here, \( R_{\scriptscriptstyle{H}} = R +  \delta + a/2 \) is the outermost contact point of the shell and electrolyte. The surface potential values are:
\begin{align}
	\varphi_{\scriptscriptstyle{H}} &= \psi(R_{\scriptscriptstyle{H}}) = \frac{\sigma_{\scriptscriptstyle{Ho}}}{\varepsilon_{\scriptscriptstyle{0}}\varepsilon \kappa}, \label{Plates-phi-H} \\
	\varphi_0 &= \psi(R +  \delta) = \varphi_{\scriptscriptstyle{H}} + \frac{a}{2} \frac{\sigma_{\scriptscriptstyle{Ho}}}{\varepsilon_{\scriptscriptstyle{0}}\varepsilon}, \label{Plates-phi-0} \\
	\psi_0 &= \psi(R) = \varphi_0 +  \delta\left( \frac{\sigma_{\scriptscriptstyle{Ho}} - \sigma_0}{\varepsilon_{\scriptscriptstyle{0}}\varepsilon} \right), \label{Plates-psi-0} \\
	\psi_{\scriptscriptstyle{H}} &= \psi(R - a/2) = \psi_0 + \frac{a}{2} \left( \frac{\sigma_{\scriptscriptstyle{Ho}} - 2\sigma_0}{\varepsilon_{\scriptscriptstyle{0}}\varepsilon} \right), \label{Plates-psi-H} \\
	\psi_{\scriptscriptstyle{d}}&\equiv\psi(r=0)=\frac{\sigma_{\scriptscriptstyle{Hi}}}{\varepsilon_{\scriptscriptstyle{0}}\varepsilon\kappa}csch[\kappa (R-a/2)].\label{Plates-psi-d}
\end{align}

\paragraph{Effective Surface Charge Density:}

The effective outer surface charge density \( \sigma_{\scriptscriptstyle{Ho}} \) is given by:
\begin{equation}
	\sigma_{\scriptscriptstyle{Ho}} = \left[ \frac{\kappa(a +  \delta) + 2 \coth[\kappa(R - a/2)]}{1 + \kappa(a +  \delta) + \coth[\kappa(R - a/2)]} \right] \sigma_0.
	\label{Plates_sigmaHO}
\end{equation}


\subsubsection{Cylindrical Shell}\label{Cylindrical-shell-EDL}

The LPB solutions for a cylindrical shell geometry (infinitely long hollow cylinder) immersed in a symmetric electrolyte, with Stern correction, are given by~\cite{Adrian-JML-2023}.

\paragraph{Mean Electrostatic Potential \(\psi(r)\):}

\begin{equation}
	A_c=
	\frac{
		R_{\scriptscriptstyle H}\sigma_{\scriptscriptstyle Ho}
		-(2R+ \delta)\sigma_0
	}
	{\varepsilon_{\scriptscriptstyle 0}\varepsilon},
\end{equation}

\begin{equation}
	B_c=
	\frac{
		R_{\scriptscriptstyle H}\sigma_{\scriptscriptstyle Ho}
		-(R+ \delta)\sigma_0
	}
	{\varepsilon_{\scriptscriptstyle 0}\varepsilon},
\end{equation}

\begin{equation}
	C_c=
	\frac{
		R_{\scriptscriptstyle H}\sigma_{\scriptscriptstyle Ho}
	}
	{\varepsilon_{\scriptscriptstyle 0}\varepsilon}.
\end{equation}

\begin{equation}
	\psi(r)=
	\begin{cases}
		\displaystyle
		-A_c
		\frac{
			I_0(\kappa r)
		}
		{
			\kappa (R-a/2)\,
			I_1[\kappa(R-a/2)]
		},
		&
		0\le r\le R-a/2,
		\\[8pt]
		
		\displaystyle
		\psi_{\scriptscriptstyle H}
		+
		A_c
		\ln\!\left(
		\frac{R-a/2}{r}
		\right),
		&
		R-a/2\le r\le R,
		\\[8pt]
		
		\displaystyle
		\psi_0
		+
		B_c
		\ln\!\left(
		\frac{R}{r}
		\right),
		&
		R\le r\le R+ \delta,
		\\[8pt]
		
		\displaystyle
		\varphi_0
		+
		C_c
		\ln\!\left(
		\frac{R+ \delta}{r}
		\right),
		&
		R+ \delta\le r\le R_{\scriptscriptstyle H},
		\\[8pt]
		
		\displaystyle
		\frac{\sigma_{\scriptscriptstyle Ho}}
		{\varepsilon_{\scriptscriptstyle 0}\varepsilon\kappa}
		\frac{
			K_0(\kappa r)
		}
		{
			K_1(\kappa R_{\scriptscriptstyle H})
		},
		&
		R_{\scriptscriptstyle H}\le r.
		\end{cases}
		\label{Cylinder-MEP}
	\end{equation}


\paragraph{Electric Field \(E(r)\):}

\begin{equation}
	E(r)=
	\begin{cases}
		\displaystyle
		A_c
		\frac{
			I_1(\kappa r)
		}
		{
			(R-a/2)\,
			I_1[\kappa(R-a/2)]
		},
		&
		0\le r\le R-a/2,
		\\[8pt]
		
		\displaystyle
		\frac{A_c}{r},
		&
		R-a/2\le r\le R,
		\\[8pt]
		
		\displaystyle
		\frac{B_c}{r},
		&
		R<r<R+ \delta,
		\\[8pt]
		
		\displaystyle
		\frac{C_c}{r},
		&
		R+ \delta\le r\le R_{\scriptscriptstyle H},
		\\[8pt]
		
		\displaystyle
		\frac{
			\sigma_{\scriptscriptstyle Ho}
		}
		{
			\varepsilon_{\scriptscriptstyle 0}
			\varepsilon
			K_1(\kappa R_{\scriptscriptstyle H})
			}
			K_1(\kappa r),
			&
			R_{\scriptscriptstyle H}\le r .
		\end{cases}
	\label{Cylinder-E-r}
\end{equation}

\paragraph{Surface Potential Values:}
\begin{align}
	\varphi_{\scriptscriptstyle{H}} &= \frac{K_0(\kappa R_{\scriptscriptstyle{H}}) \sigma_{\scriptscriptstyle{Ho}}}{\varepsilon_{\scriptscriptstyle{0}}\varepsilon \kappa K_1(\kappa R_{\scriptscriptstyle{H}})}, \label{Cyl-phi-H} \\
	\varphi_0 &= \varphi_{\scriptscriptstyle{H}} + \frac{R_{\scriptscriptstyle{H}} \sigma_{\scriptscriptstyle{Ho}}}{\varepsilon_{\scriptscriptstyle{0}}\varepsilon} \ln\left( \frac{R_{\scriptscriptstyle{H}}}{R +  \delta} \right), \label{Cyl-phi-0} \\
	\psi_0 &= \varphi_0 + \left( \frac{R_{\scriptscriptstyle{H}} \sigma_{\scriptscriptstyle{Ho}} - (R +  \delta)\sigma_0}{\varepsilon_{\scriptscriptstyle{0}}\varepsilon} \right) \ln\left( \frac{R +  \delta}{R} \right), \label{Cyl-psi-0} \\
	\psi_{\scriptscriptstyle{H}} &= \psi_0 + \left( \frac{R_{\scriptscriptstyle{H}} \sigma_{\scriptscriptstyle{Ho}} - (2R +  \delta)\sigma_0}{\varepsilon_{\scriptscriptstyle{0}}\varepsilon} \right) \ln\left( \frac{R}{R - \frac{a}{2}} \right), \label{Cyl-psi-H} \\
	\\ 	\psi_{\scriptscriptstyle{d}} &= \psi_{\scriptscriptstyle{H}} + \frac{\left[R_{\scriptscriptstyle{H}}\sigma_{\scriptscriptstyle{Ho}}-(2R+ \delta)\sigma_{o}\right](I_{\scriptscriptstyle{0}}[\kappa(R-a/2)]-1)}{\varepsilon_{\scriptscriptstyle{0}}\varepsilon\kappa(R-a/2)I_{\scriptscriptstyle{1}}[\kappa(R-a/2)]}. \label{Cyl-psi-d}
\end{align}

\paragraph{Effective Surface Charge:}
\begin{equation}
	\sigma_{\scriptscriptstyle{Ho}} = \frac{L^{cyl}_2}{L^{cyl}_1} \sigma_0
	\label{Cylinder-sigmaHo}
\end{equation}
where:
\begin{equation}\label{L1-Cyl}
		\begin{split}
		L^{cyl}_1={}
		\frac{
			R_H I_0\big(\kappa(R-a/2)\big)
		}
		{
			(R-a/2)\kappa I_1\big(\kappa(R-a/2)\big)
		}\\
		+
		R_H\ln\bigg(\frac{R_H}{R-a/2}\bigg)
		+
		\frac{
			K_0(\kappa R_H)
		}
		{
			\kappa K_1(\kappa R_H)
		},
			\end{split}
\end{equation}

\begin{equation}\label{L2-Cyl}
	\begin{split}
		L^{cyl}_2=&{}
		\frac{
			(2R+\delta) I_0\big(\kappa(R-a/2)\big)
		}
		{
			(R-a/2)\kappa I_1\big(\kappa(R-a/2)\big)
		}\\
		&+
		(2R+\delta)\ln\bigg(\frac{R}{R-a/2}\bigg)\\
		&+
		(R+\delta)\ln\bigg(\frac{R+\delta}{R}\bigg).
	\end{split}
\end{equation}

\subsubsection{Spherical Shell}\label{Spherical-shell-EDL}

The LPB analytical solution for a spherical shell immersed in a symmetric electrolyte, with Stern correction, is given by~\cite{Adrian-JML-2023}.

\paragraph{Mean Electrostatic Potential \(\psi(r)\):}

Here, the denominator
\[
D(R) = \sinh[\kappa (R - \frac{a}{2})] - \kappa (R - \frac{a}{2}) \cosh[\kappa (R - \frac{a}{2})]
\]

\begin{equation}
	A_s=
	\frac{
		R_{\scriptscriptstyle H}^2\sigma_{\scriptscriptstyle Ho}
		-\big[(R+ \delta)^2+R^2\big]\sigma_0
	}
	{\varepsilon_{\scriptscriptstyle 0}\varepsilon},
\end{equation}

\begin{equation}
	B_s=
	\frac{
		R_{\scriptscriptstyle H}^2\sigma_{\scriptscriptstyle Ho}
		-(R+ \delta)^2\sigma_0
	}
	{\varepsilon_{\scriptscriptstyle 0}\varepsilon},
	\qquad
	C_s=
	\frac{
		R_{\scriptscriptstyle H}^2\sigma_{\scriptscriptstyle Ho}
	}
	{\varepsilon_{\scriptscriptstyle 0}\varepsilon}.
\end{equation}

\begin{equation}
	\psi(r)=
	\begin{cases}
		\displaystyle
		\frac{A_s}{D(R)}
		\frac{\sinh(\kappa r)}{r},
		&
		0\le r\le R-a/2,
		\\[8pt]
		
		\displaystyle
		\psi_{\scriptscriptstyle H}
		+
		\frac{A_s}{R-a/2}
		\left(
		\frac{R-a/2}{r}-1
		\right),
		&
		R-a/2\le r\le R,
		\\[8pt]
		
		\displaystyle
		\psi_0
		+
		\frac{B_s}{R}
		\left(
		\frac{R}{r}-1
		\right),
		&
		R\le r\le R+ \delta,
		\\[8pt]
		
		\displaystyle
		\varphi_0
		+
		\frac{C_s}{R+ \delta}
		\left(
		\frac{R+ \delta}{r}-1
		\right),
		&
		R+ \delta\le r\le R_{\scriptscriptstyle H},
		\\[8pt]
		
		\displaystyle
		\frac{C_s}{1+\kappa R_{\scriptscriptstyle H}}
		\frac{
			e^{-\kappa(r-R_{\scriptscriptstyle H})}
		}{r},
		&
		R_{\scriptscriptstyle H}\le r .
		\end{cases}
		\label{Sphere-MEP}
	\end{equation}
	
	Here, the denominator
	\[
	D(R) = \sinh[\kappa (R - \frac{a}{2})] - \kappa (R - \frac{a}{2}) \cosh[\kappa (R - \frac{a}{2})]
	\]

\paragraph{Electric Field \(E(r)\):}

\begin{equation}
	E(r)=
	\begin{cases}
		\displaystyle
		\frac{A_s}{D(R)}
		\frac{
			\sinh(\kappa r)-\kappa r\cosh(\kappa r)
		}
		{r^2},
		&
		0\le r\le R-a/2,
		\\[8pt]
		
		\displaystyle
		\frac{A_s}{r^2},
		&
		R-a/2\le r\le R,
		\\[8pt]
		
		\displaystyle
		\frac{B_s}{r^2},
		&
		R<r<R+ \delta,
		\\[8pt]
		
		\displaystyle
		\frac{C_s}{r^2},
		&
		R+ \delta\le r\le R_{\scriptscriptstyle H},
		\\[8pt]
		
		\displaystyle
		\frac{C_s(1+\kappa r)}
		{1+\kappa R_{\scriptscriptstyle H}}
		\frac{
			e^{-\kappa(r-R_{\scriptscriptstyle H})}
		}
		{r^2},
		&
		R_{\scriptscriptstyle H}\le r .
	\end{cases}
	\label{Sphere-E-r}
\end{equation}

\paragraph{Surface Potential Values:}

\begin{align}
	\varphi_{\scriptscriptstyle{H}} &= \frac{R_{\scriptscriptstyle{H}} \sigma_{\scriptscriptstyle{Ho}}}{\varepsilon_{\scriptscriptstyle{0}}\varepsilon (1 + \kappa R_{\scriptscriptstyle{H}})}, \label{Sph-phi-H} \\
	\varphi_0 &= \varphi_{\scriptscriptstyle{H}} + \frac{R_{\scriptscriptstyle{H}} \sigma_{\scriptscriptstyle{Ho}}}{\varepsilon_{\scriptscriptstyle{0}}\varepsilon (R +  \delta)} \cdot \frac{a}{2}, \label{Sph-phi-0} \\
	\psi_0 &= \varphi_0 + \frac{[R_{\scriptscriptstyle{H}}^2 \sigma_{\scriptscriptstyle{Ho}} - (R +  \delta)^2 \sigma_0]  \delta}{\varepsilon_{\scriptscriptstyle{0}}\varepsilon R (R +  \delta)}, \label{Sph-psi-0} \\
	\psi_{\scriptscriptstyle{H}} &= \psi_0 + \frac{[R_{\scriptscriptstyle{H}}^2 \sigma_{\scriptscriptstyle{Ho}} - (R+ \delta)^2 - R^2]\sigma_0 \cdot \frac{a}{2}}{\varepsilon_{\scriptscriptstyle{0}}\varepsilon R(R - \frac{a}{2})}, \label{Sph-psi-H} \\ \psi_{\scriptscriptstyle{d}}& =\\& \frac{\kappa}{\varepsilon_{\scriptscriptstyle{0}}\varepsilon} \left\{ \frac{[R_{\scriptscriptstyle{H}}^2\sigma_{\scriptscriptstyle{Ho}}-\left[ R^2 + (R+ \delta)^2 \right]\sigma_{\scriptscriptstyle{0}}]}{\sinh[\kappa (R-a/2)]-\kappa (R-a/2)\cosh[\kappa(R-a/2)]}\right\}.\label{Sph-psi-d}	
\end{align}

\paragraph{Effective Surface Charge:}
\begin{equation}
	\sigma_{\scriptscriptstyle{Ho}} = \frac{L^{sph}_2}{L^{sph}_1} \sigma_0
	\label{Sphere-sigmaHo}
\end{equation}

where:

\begin{equation}\label{L1-Sph}
	\begin{split}
		L^{sph}_1 &= \frac{R_{\scriptscriptstyle{H}}}{1 + \kappa R_{\scriptscriptstyle{H}}} + \frac{\frac{a}{2} R_{\scriptscriptstyle{H}}}{R + \delta} + \frac{\delta R_{\scriptscriptstyle{H}}^2}{R(R + \delta)} +\\
		& \frac{\frac{a}{2} R_{\scriptscriptstyle{H}}^2}{R(R - \frac{a}{2})} 
		- \frac{\sinh[\kappa(R - \frac{a}{2})] R_{\scriptscriptstyle{H}}^2}{(R - \frac{a}{2}) D(R)} \\[5pt]
	\end{split}
\end{equation}
\begin{equation}\label{L2-Sph}
	\begin{split}
		&L^{sph}_2 = \frac{d(R + \delta)}{R} + \frac{\frac{a}{2} [(R + \delta)^2 + R^2]}{R(R - \frac{a}{2})} \\
		&- \frac{\sinh[\kappa(R - \frac{a}{2})] [(R + \delta)^2 + R^2]}{(R - \frac{a}{2}) D(R)}
	\end{split}
\end{equation}

\section{Recovery of local electroneutrality in the limit $R\to\infty$}
\label{appendix-B}

\setcounter{equation}{0}

In this appendix we show that local electroneutrality is recovered in the
limit \(R\to\infty\) for spherical and cylindrical nanocavities. We use
\(\delta\) for the shell-wall thickness and define
\begin{equation}
	R_H=R+\delta+\frac{a}{2}.
\end{equation}

\subsection{Cylindrical geometry}

For the cylindrical shell we use the following asymptotic properties of
modified Bessel functions:
\begin{equation}
	\lim_{r\to\infty}
	\frac{I_\nu(r)}{e^r/\sqrt{2\pi r}}
	=1,
	\qquad
	\lim_{r\to\infty}
	\frac{K_\nu(r)}{\sqrt{\pi/(2r)}\,e^{-r}}
	=1 .
\end{equation}

Therefore,
\begin{equation}
	\lim_{R\to\infty}
	\frac{
		I_0\big(\kappa(R-a/2)\big)
	}
	{
		I_1\big(\kappa(R-a/2)\big)
	}
	=1,
	\qquad
	\lim_{R\to\infty}
	\frac{
		K_0(\kappa R_H)
	}
	{
		K_1(\kappa R_H)
	}
	=1 .
\end{equation}

We also use the elementary limit
\begin{equation}
	\lim_{R\to\infty}
	(R+C)\ln\bigg(\frac{R+A}{R+B}\bigg)
	=
	A-B,
\end{equation}
where \(A\), \(B\), and \(C\) are real constants.

For the cylindrical shell,
\begin{equation}
	\sigma_{\rm Ho}=\frac{L^{cyl}_2}{L^{cyl}_1}\sigma_0 ,
\end{equation}

From \cref{L1-Cyl,L2-Cyl} and using the limits above, one obtains
\begin{equation}
	\lim_{R\to\infty}L^{cyl}_1
	=
	\frac{1}{\kappa}
	+\delta+a
	+\frac{1}{\kappa}
	=
	\frac{2}{\kappa}+\delta+a ,
\end{equation}
and
\begin{equation}
	\lim_{R\to\infty}L^{cyl}_2
	=
	\frac{2}{\kappa}
	+\frac{a}{2}
	+\frac{a}{2}
	+\delta
	=
	\frac{2}{\kappa}+\delta+a .
\end{equation}

Hence,
\begin{equation}
	\lim_{R\to\infty}\frac{L^{cyl}_2}{L^{cyl}_1}=1,
\end{equation}
and therefore
\begin{equation}
	\lim_{R\to\infty}\sigma_{\rm Ho}=\sigma_0 .
\end{equation}

For the cylindrical geometry,
\begin{equation}
	\sigma_{\rm Hi}
	=
	\frac{
		(R+\delta+a/2)\sigma_{\rm Ho}
		-
		(2R+\delta)\sigma_0
	}
	{R-a/2}.
\end{equation}

Taking the limit \(R\to\infty\),
\begin{equation}
	\lim_{R\to\infty}\sigma_{\rm Hi}
	=
	\sigma_0-2\sigma_0
	=
	-\sigma_0 .
\end{equation}

Since
\begin{equation}
	E(R-a/2)=\frac{\sigma_{\rm Hi}}{\epsilon_0\epsilon},
\end{equation}
it follows that
\begin{equation}
	\lim_{R\to\infty}E(R-a/2)
	=
	-\frac{\sigma_0}{\epsilon_0\epsilon}.
\end{equation}

Therefore, if
\begin{equation}
	E(R)=\frac{\sigma_0}{\epsilon_0\epsilon},
\end{equation}
then
\begin{equation}
	\lim_{R\to\infty}
	\big[
	E(R-a/2)+E(R)
	\big]
	=0 .
\end{equation}

\subsection{Spherical geometry}

For the spherical shell, the surface charge density at the outer surface can
be written as
\begin{equation}
	\sigma_{\rm Ho}=\frac{L^{sph}_2}{L^{sph}_1}\sigma_0 .
\end{equation}

The quantities \(L^{sph}_1\) and \(L^{sph}_2\) are


It is useful to write the ratio as
\begin{equation}
	\frac{L^{sph}_2}{L^{sph}_1}=f_1f_2f_3,
\end{equation}
where $L^{sph}_1$ and $L^{sph}_2$ are given by \cref{L1-Sph,L2-Sph}, and
\begin{equation}
	f_1=
	\frac{2(1+\kappa R_H)}{4R_H\kappa},
\end{equation}
\begin{equation}
	\begin{split}
		f_2
		&=
		\frac{
			4R+2\delta
		}
		{R+\delta+a/2}+\\
		&
		\frac{
			\kappa
			\big[
			2\delta(R+\delta)+a(2R+\delta)
			\big]
			\coth\big(\kappa(R-a/2)\big)
		}
		{R+\delta+a/2}.\\
		&\\
		&
	\end{split}
\end{equation}
and
\begin{equation}
	f_3=
	\frac{1}
	{
		1+
		\big[1+\kappa(\delta+a)\big]
		\coth\big(\kappa(R-a/2)\big)
	}.
\end{equation}

Since
\begin{equation}
	\lim_{R\to\infty}f_1=\frac12,
\end{equation}
\begin{equation}
	\lim_{R\to\infty}f_2=4+2\kappa(\delta+a),
\end{equation}
and
\begin{equation}
	\lim_{R\to\infty}f_3=
	\frac{1}{2+\kappa(\delta+a)},
\end{equation}
we find
\begin{equation}
	\lim_{R\to\infty}\frac{L^{sph}_2}{L^{sph}_1}
	=
	\frac12
	\big[4+2\kappa(\delta+a)\big]
	\frac{1}{2+\kappa(\delta+a)}
	=1 .
\end{equation}

Therefore,
\begin{equation}
	\lim_{R\to\infty}\sigma_{\rm Ho}
	=
	\lim_{R\to\infty}
	\frac{L^{sph}_2}{L^{sph}_1}\sigma_0
	=
	\sigma_0 .
\end{equation}

For the spherical geometry,
\begin{equation}
	\sigma_{\rm Hi}
	=
	\frac{
		(R+\delta+a/2)^2\sigma_{\rm Ho}
		-
		\big[(R+\delta)^2+R^2\big]\sigma_0
	}
	{(R-a/2)^2}.
\end{equation}

Thus,
\begin{equation}
	\lim_{R\to\infty}\sigma_{\rm Hi}
	=
	\sigma_0-2\sigma_0
	=
	-\sigma_0 .
\end{equation}

Since
\begin{equation}
	E(R-a/2)=\frac{\sigma_{\rm Hi}}{\epsilon_0\epsilon},
\end{equation}
it follows that
\begin{equation}
	\lim_{R\to\infty}E(R-a/2)
	=
	-\frac{\sigma_0}{\epsilon_0\epsilon}.
\end{equation}

Therefore, if
\begin{equation}
	E(R)=\frac{\sigma_0}{\epsilon_0\epsilon},
\end{equation}
then
\begin{equation}
	\lim_{R\to\infty}
	\big[
	E(R-a/2)+E(R)
	\big]
	=0 .
\end{equation}


%

\end{document}